\documentclass[prd,onecolumn,nofootinbib,floatfix,superscriptaddress]{revtex4}
\usepackage{amsmath, amssymb, graphicx, comment, amsthm, relsize}
\usepackage{bbold, bm}
\usepackage{subfigure}
\usepackage[nohug]{diagrams}
\usepackage{enumerate}

\theoremstyle{definition}

\theoremstyle{remark}

\newcommand{\be}{\begin{eqnarray}}
\newcommand{\ee}{\end{eqnarray}}

\newcommand{\bcf}{\begin{figure}}
\newcommand{\ecf}{\end{figure}}

\newcommand{\SU}{\text{SU}}

\newcommand{\su}{\mathfrak{su}}
\newcommand{\so}{\mathfrak{so}}
\newcommand{\nnn}{\nonumber\\}

\def\Tr{\text{Tr}}
\def\Pexp{\overrightarrow{\exp}}
\def\A{\bm{A}}
\def\e{\bm{e}}
\def\X{\bm{X}}

\def\F{\bm{F}}

\def\l{\ell}

\def\t{\bm{\tau}}
\def\n{\mathsf{n}}

\def\v{\mathsf{v}}

\def\ed{\mathsf{e}}
\def\rd{\mathrm{d}}

\def\t{\bm{\tau}}

\def\sslash{/ \! \! /}
\def\z{\bm{z}}

\def\r{\bm{r}}

\def\t{\bm{\tau}}
\def\b{\bm{b}}

\def\u{\bm{u}}

\def\q{\bm{q}}
\def\p{\bm{p}}
\def\L{\bm{\lambda}}
\def\bP{\bm{P}}

\def\x{\hat{\bm{x}}}
\def\y{\hat{\bm{y}}}

\def\B{\mathcal{B}}
\def\N{\bm{N}}
\def\r{\mathsf{r}}

\def\ti{\hat{\bm{t}}}

\def\o{\mathsf{o}}

\begin{document}

\title{Point particles in 2+1 dimensions:\\
       general relativity and loop gravity descriptions}
\author{Jonathan Ziprick}
\email{jziprick@unb.ca}
\affiliation{University of New Brunswick\\
Department of Mathematics and Statistics\\
Fredericton, NB E3B 5A3, Canada}
\date{\today}

\begin{abstract}
We develop a Hamiltonian description of point particles in (2+1)-dimensions using connection and frame-field variables for general relativity. The topology of each spatial hypersurface is that of a punctured two-sphere with particles residing at the punctures. We describe this topology with a CW complex (a collection of two-cells glued together along edges), and use this to fix a gauge and reduce the Hamiltonian. The equations of motion for the fields describe a dynamical triangulation where each vertex moves according to the equation of motion for a free relativistic particle. The evolution is continuous except for when triangles collapse (i.e. the edges become parallel) causing discrete, topological changes in the underlying CW complex.

We then introduce the loop gravity phase space parameterized by holonomy-flux variables on a graph (a network of one-dimensional links). By embedding a graph within the CW complex, we find a description of this system in terms of loop variables. The resulting equations of motion describe the same dynamical triangulation as the connection and frame-field variables. In this framework, the collapse of a triangle causes a discrete change in the underlying graph, giving a concrete realization of the graph-changing moves that many expect to feature in full loop quantum gravity. The main result is a dynamical model of loop gravity which agrees with general relativity and is well-suited for quantization using existing methods.
\end{abstract}

\maketitle

\section{Introduction}

Three-dimensional gravity has often served as a useful toy model for the full 4d theory. Without matter, the model is very simple since it does not possess any local degrees of freedom. A more interesting model can be obtained by including a number of point particles. These give rise to a finite number of physical degrees of freedom associated to the particle positions and momenta.

The study of point particles in 3d gravity has a rich history, with the first static solutions for one- and two-body models being found in 1963 by Staruszkiewicz \cite{Star}. Interest in the field really began to grow around twenty years later after Deser, Jackiw and 't Hooft (DJH) published their seminal work on the subject \cite{DJH}. This article helped to clarify the picture, giving a simple representation of dynamical point particles as conical singularities moving in space. Each singularity is described within a Minkowski spacetime by cutting out a `wedge' and identifying the sides of the wedge. These identifications encode the deficit angle of a cone, which represents the mass of the particle. Later studies \cite{tHooft1,Kadar,Matschull} have built upon this idea, using a partition of two-dimensional spatial hypersurfaces into flat polygons where particles reside at vertices. In this setup, deficit angles are encoded in the matching conditions used to `glue' the polygons together along edges, and dynamics are seen as changes in the length and orientation of the edges. These studies resemble the approach we take here.

Beyond classical gravity, point particles in three dimensions also make a useful model for the study of quantum gravity since the theory can be solved explicitly. This can be done in a number of different ways, and the book \cite{Carlip-book} by Steve Carlip provides a nice overview. We note in particular that this model has been solved using spin foams \cite{FL}, which is closely related to loop quantum gravity (LQG). Spin foams are a path integral (Lagrangian) approach to quantum gravity, while LQG is the canonical (Hamiltonian) version of the theory. In 3d, the spin foam approach is built upon a decomposition of spacetime into polyhedron-shaped cells so that each spatial slice is a two-space composed of polygons that are glued together along their boundaries. The canonical approach is therefore well-suited for describing the evolution of point particles in 3d gravity in terms of dynamical polygons, similarly to the studies mentioned above.

\begin{figure}
\begin{center}
\includegraphics[width=0.3\linewidth]{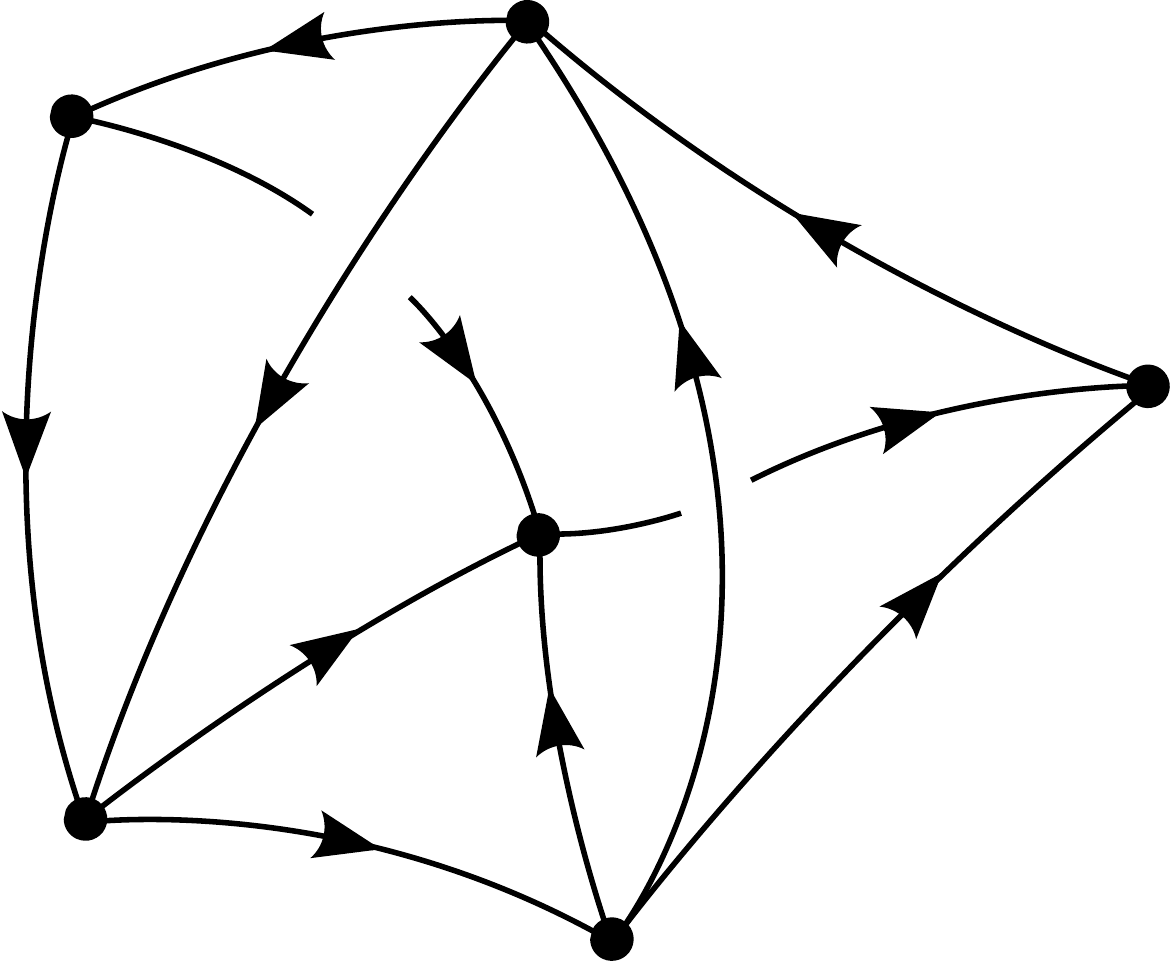}
\caption[Example of a graph]{\label{graph}An example of a three-dimensional graph. Each link is drawn with an orientation, and intersections between links occur only at the nodes.}
\end{center}
\end{figure}

The full Hilbert space $\mathcal{H}$ of LQG is constructed from a family of finite-dimensional spin network Hilbert spaces $\mathcal{H}_\Gamma$, each defined upon an oriented graph $\Gamma$. For our purposes, an oriented graph (or simply `graph' hereafter) is a topological space composed of one-dimensional links, each possessing an orientation, which are joined with other links at their endpoints. See fig. \ref{graph} for an example. Loop quantum gravity is a continuous theory, and the full infinite-dimensional Hilbert space $\mathcal{H}$ is obtained in the \textit{projective} limit \cite{projective}, which takes into account the infinite number of spin network spaces $H_\Gamma$ associated to a certain sequence of graphs $\Gamma$. Now, underlying this quantum theory is a classical theory constructed in an analogous manner, where the full phase space $\mathcal{P}$ is obtained in the projective limit \cite{QSD7} of spin network phase spaces $P_\Gamma$ associated to graphs $\Gamma$. In other words, the states and operators of LQG can be constructed from these phase spaces. 

In the LQG framework, we are still unable to formulate the physical observables of a gravitational theory, and it has been particularly difficult to develop a dynamics which is consistent with general relativity in the appropriate limits. We hope to shed some light on these issues by looking more closely at the classical theory of loop gravity in terms of phase spaces. Although we have recently developed an understanding of the kinematics \cite{FGZ,FZ} and how this relates to spatial geometries in general relativity, loop classical gravity remains largely unexplored. In this article we work toward understanding general relativity in terms of the loop variables by studying a simple gravitational model.

In this paper we consider 3d gravity coupled to a number of spinless point particles\footnote{See \cite{Z1} for a precursor to this work.}. We begin with a canonical decomposition of the first order action for pure gravity, written in terms of a connection $\A$ and a frame-field $\e$. Once we have established the pure gravity Hamiltonian, we introduce particles by excising a set of points from the space manifold and introducing constraints which fix the particle masses. After a careful analysis of boundary terms and conditions, we perform an explicit gauge fix by first defining the spatial topology as a CW complex (a collection of two-cells glued together), which leaves only a finite number of degrees of freedom associated to the particles.

The gauge-fixed fields $(\A, \e)$ describe a geometry where each cell of the CW complex is a triangle, and the equations of motion for these fields imply that the vertices of these triangles are dynamical. Each triangle changes shape according to the motion of its vertices, and it can happen that one or more triangles collapse when a vertex meets an edge or another vertex. This invokes a discrete change in the CW complex and an associated definition of the fields $(\A, \e)$ in the affected triangles. After this discrete change, the evolution is again continuous until another triangle collapse.

After establishing the description in terms of field variables for general relativity, we then map to the loop variables and develop an alternative description. The first step is to embed a graph $\Gamma$ within the CW complex. This allows us to map from the fields $(\A, \e)$ to a discrete set of variables on a graph $\Gamma$ which parameterize the phase space $P_\Gamma$. We find that the loop gravity variables describe the same triangulation as the fields $(\A, \e)$, and have equivalent equations of motion. In this setting, the collapse of a triangle signals a discrete change in the underlying graph. The loop gravity description continues to agree with the relativistic description throughout this evolution. The result is that we have two equivalent descriptions of point particles in 3d gravity: the relativistic version in terms of fields $(\A, \e)$ on a CW complex, and the loop gravity version in terms of the loop variables on a graph.

Having a complete canonical description of this model in terms of loop variables, we are able to calculate observables and formulate a dynamics that is consistent with general relativity. The CW complex plays an important role as the topology underlying the field variables for general relativity, and this structure helps us translate into the loop gravity framework which uses a discrete set of variables on a graph. We give a new perspective on how the dynamics of LQG may play out, with continuous evolution of the loop variables on top of a topological graph, and this graph may undergo discrete changes. These discrete graph moves are a concrete realization of the graph changes expected to feature in LQG in order to describe propagating gravitons \cite{Smolin}. All of this goes toward setting up a quantization via the well-established framework of LQG, making an interesting toy model for the full four-dimensional theory.

\section{Hamiltonian analysis}
The first order formalism of general relativity parameterizes the gravitational field in terms of a connection $\A$ and a frame-field $\e$. We write the spacetime manifold as $M=\mathbb{R} \times \Sigma$ where $\Sigma$ is a spacelike surface homeomorphic to $S^2$. In a 4d Lorentzian spacetime, the field variables defined upon spatial hypersurfaces take values in the $\su(2)$ algebra \cite{Ashtekar}. In this three-dimensional setting, we choose to work with a Riemannian rather than a Lorentzian spacetime since the field variables are then $\su(2)$-valued as in the 4d Lorentzian case\footnote{In a 3d Lorentzian spacetime the field variables take values in the $\so(1,2)$ algebra.}, providing a theory that bears more similarity to the case of full gravity.

Before we begin, let us briefly introduce some useful notation. We use $\su(2)$ basis elements ${\bm \tau}^i$ (for $i=0,1,2$) which are given by $-i/2$ times the Pauli matrices. With this choice, the basis elements satisfy $\Tr(\t^i \t^j) = -\frac{1}{2}\delta^{ij}$ and have a commutator bracket given by $[\t^i, \t^j] = \epsilon^{ijk} \t^k$. When the entries of this bracket are form-fields, then $[\cdot, \cdot]$ implies taking the $\su(2)$ commutator \textit{and} the wedge product between the elements within the bracket. We shall write elements $\bm{v} \in \su(2)$ in bold font where $\bm{v} \equiv v^i {\bm \tau}^i$, and internal indices are written `up' or `down' according to convenience (since they are `raised' and `lowered' by Kronecker deltas). An element of $\bm{v} \in \su(2)$ has three components and can be seen as a vector in $\mathbb{R}^3$ where the trace plays the role of a dot product and the commutator bracket is the cross product. We shall refer to the modulus or norm as $v^2 \equiv -2 \Tr (\bm{v} \bm{v}) = v^i v^i$, and the direction as $v^i / v$. Where coordinate indices are explicitly shown, we shall use Greek letters $\mu = 0,1,2$ to label spacetime indices and Latin letters $a = 1, 2$ to label space indices. Finally, we work in units such that $8 \pi G = c = 1$.

\subsection{Pure gravity}
We begin with an analysis of pure gravity before bringing matter into the picture since the singular nature of point particles requires some special treatment.
The first order action for pure gravity is:
\be
S = -\frac{1}{2} \int_M \rd^3\!x \epsilon^{\mu \nu \rho} e_\mu^i F^i_{\nu \rho},
\ee
where $F^i_{\mu \nu} = \partial_\mu A_\nu^i - \partial_\nu A_\mu^i + \epsilon^{ijk}A^j_\mu A^k_\nu$ is the curvature of the connection.
Writing the spacetime indices as $\mu = 0, a$, the action is decomposed into time and space components as follows:
\be
S &=& -\frac{1}{2} \int dt \int_\Sigma \rd^2\!x \left( \epsilon^{0ab} e_0^i F^i_{ab} + 2\epsilon^{ab0} e_a^i F^i_{b0} \right) \nnn
&=& -\frac{1}{2} \int dt \int_\Sigma \rd^2\!x \epsilon^{ab} \left[ e_0^i F^i_{ab} + 2e_a^i \left( \partial_b A_0^i - \partial_0 A_b^i + \epsilon^{ijk}A^j_b A^k_0 \right) \right] \nnn
&=& -\int dt \int_\Sigma \rd^2\!x \epsilon^{ab} \left[ e_b^i \dot{A}_a^i + \frac{1}{2} e_0^i F^i_{ab} + A_0^i \left( \partial_a e_b^i + \epsilon^{ijk}A^j_a e_b^k \right)\right] \nnn
\label{action}
&=& \int dt \int_\Sigma \rd^2\!x \epsilon^{ab} \left[ e_a^i \dot{A}_b^i - N^i F^i_{ab} - \lambda^i G^i_{ab} \right]
\ee
where the over-dot denotes a derivative with respect to the arbitrary time parameter $t$, $\epsilon^{\mu \nu \rho}$ is the completely antisymmetric, metric-independent tensor density and $\epsilon^{ab} \equiv \epsilon^{0ab}$. The time components of the fields are Lagrange multipliers, written in the last line as $N^i:= \frac{1}{2}e_0^i$ and $\lambda^i:= \frac{1}{2}A_0^i$ corresponding respectively to the flatness constraint:
\be
\label{flatness}
F^i_{ab} := \partial_a A^i_b - \partial_b A^i_a + \epsilon^{ijk} A^j_a A^k_b,
\ee
and the Gauss constraint:
\be
\label{Gauss}
G^i_{ab}:= \partial_a e_b^i - \partial_b e_a^i + \epsilon^{ijk}(A^j_a e_b^k - A^j_b e_a^k ).
\ee
Notice that the Gauss constraint in 3d is equivalent to a zero-torsion constraint.

We choose $A_a^i$ as the configuration variable and find that the canonical momentum is $e_a^i$. The Poisson brackets are:
\be
\label{AePB}
\left\{ A^i_a(x) , A^j_b(y)\right\} = \left\{ e_a^i(x) , e_b^j(y) \right\} =0 , \qquad \qquad \left\{ A^i_a(x) , e_b^j(y) \right\} = \epsilon_{ab} \delta^{ij} \delta^2(x-y) .
\ee
These variables parameterize a continuous phase space $(\A, \e) \in \mathcal{P}$ for general relativity. There are $2 \times 6 = 12$ degrees of freedom per point, and the two constraints completely constrain these variables leaving no degrees of freedom in the case of pure gravity.

We write the smeared constraints as:
\be
\mathcal{F}(\bm{N}) := \int_\Sigma \rd^2\!x \epsilon^{ab}N^i F^i_{ab} = \int_\Sigma N^i F^i, \\
\mathcal{G}(\bm{\lambda}) := \int_\Sigma \rd^2\!x \epsilon^{ab}\lambda^i G^i_{ab} = \int_\Sigma \lambda^i G^i,
\ee
where we have used differential form notation to write the curvature and torsion without coordinate indices, e.g. $F^i \equiv F^i_{ab} \rd x^a \wedge \rd x^b$. If we also use our notation for elements of $\su(2)$, e.g. $\bm{F} \equiv F^i \tau^i$, the curvature is written simply as $\bm{F} = \rd_A \bm{A} = \rd \bm{A} + \frac{1}{2}[\bm{A},\bm{A}]$, and the torsion as $\bm{G} = \rd_A \bm{e} = \rd \bm{e} + [\bm{A},\bm{e}]$. The flatness and Gauss constraints have the same form in this theory, putting the frame-field and connection on the same footing.
The constraints form a first class algebra:
\be
\label{PBs}
\left\{\mathcal{F}(\bm{N}), \mathcal{F}(\tilde{\bm{N}}) \right\} &=& 0 , \nnn
\left\{\mathcal{F}(\bm{N}), \mathcal{G}(\bm{\lambda}) \right\} &=& \mathcal{F}(]\bm{N},\bm{\lambda}]), \nnn
\left\{\mathcal{G}(\bm{\lambda}), \mathcal{G}(\tilde{\bm{\lambda}}) \right\} &=& \mathcal{G} ([\bm{\lambda}, \tilde{\bm{\lambda}}]),
\ee

The Gauss constraint generates $\SU(2)$-gauge transformations. The infinitesimal transformations are:
\be
\delta_{\mathcal{G}} A^i = \left\{ A^i, \mathcal{G}(\bm{\lambda}) \right\} = -\rd_A\lambda^i , \hspace{1in} \delta_{\mathcal{G}} e^i = \left\{ e^i, \mathcal{G}(\bm{\lambda}) \right\} = \epsilon^{ijk} e^j \lambda^k .
\ee
For a function $g(x) \in \SU(2)$, the finite transformations are:
\be
\bm{A} \rightarrow g \bm{A} g^{-1} + g \rd g^{-1}, \hspace{1in} \bm{e} \rightarrow g \bm{e} g^{-1} .
\ee
These are the Riemannian analogs of Lorentz boosts and rotations.

The flatness constraint generates the following infinitesimal transformations:
\be
\delta_{\mathcal{F}} A^i = \left\{ A^i, \mathcal{F}(\bm{N}) \right\} = 0 , \hspace{1in} \delta_{\mathcal{F}} e^i = \left\{ e^i, \mathcal{F}(\bm{N}) \right\} = \rd_A N^i .
\ee
Notice these transformations do not affect the connection. For a function $\phi(x) \in \su(2)$, the finite gauge transformations are:
\be
\bm{A} \rightarrow \bm{A}, \hspace{1in} \bm{e} \rightarrow \bm{e} + \rd_A \bm{\phi} .
\ee
This shift of the triad is equivalent to a translation.

From the action (\ref{action}) we read off the Hamiltonian to be a sum of constraints:
\be
\label{ham}
H_0 = \mathcal{F}(\bm{N}) + \mathcal{G}(\bm{\lambda}).
\ee
This is the Hamiltonian for pure 3d Riemannian gravity on $S^2 \times \mathbb{R}$ parameterized by a connection and frame-field $(\bm{A}, \bm{e}) \in \mathcal{P}$.

\subsection{Gravity with particles}
Let us now bring a number $|\v|$ of point particles into the picture. It is well-known that point particles in a 3d spacetime manifest themselves as conical singularities \cite{DJH}, so that defining the field variables at these locations is problematic. We handle this problem by excising the particle worldlines from the spacetime, similarly to the methods used in \cite{Matschull}. Excising these worldlines corresponds to treating each particle location on a constant-time hypersurface $\Sigma_t$ as a puncture. Now, having removed these points from the spatial manifold, we will need to find the particle degrees of freedom within the fields $(\bm{A},\bm{e})$ outside of the particle locations. This will be done by defining integrals of the fields near the punctures. In particular, we shall need to integrate the connection along loops which encircle the punctures, and these integrals must remain well-defined even as the loop shrinks to zero radius. For this purpose we introduce a particle boundary $\B_\v$ around each particle defined as follows. In the neighborhood of a single particle $\v$, consider a circle which goes around this particle, parameterized by a radius $r$ and an angle $\phi$. The boundary $\B_\v$ associated to the particle is taken to be this circle in the limit of vanishing radius $r \rightarrow 0$. In this limit we take $\B_\v$ to maintain its $S^1$ nature (rather than shrinking to a point) so that we can integrate around it in a well-defined manner. This is done by distinguishing points with different $\phi$ values in the $r \rightarrow 0$ limit; the only identification we make is $\phi = \phi + 2 \pi$. We associate such a boundary $\B_\v$ to each particle on $\Sigma_t$, making each spatial hypersurface homeomorphic to the punctured two-sphere $S^2 \setminus \{ \v \}$. The boundary of $\Sigma_t$ is the disjoint union of a boundary $\B_\v$ for each particle:
\be
\partial \Sigma_t = \underset{\v}{\sqcup} \B_\v.
\ee
As is commonly done, we shall drop the subscript $t$ labeling spatial hypersurfaces where it is not relevant.

In order that a particle boundary $\B_\v$ appears point-like in a hypersurface $\Sigma$, the component of the frame-field that is tangent to $\B_\v$ must be zero \cite{Matschull} so that its circumference vanishes. We can use a variable $s=[0,1]$ to parameterize the loop $\B_\v(s)$ such that a vector tangent to the boundary is given by $(\dot{\B}_\v)^a$, where the over-dot represents a derivative with respect to $s$. We ensure that each particle boundary appears point-like by introducing a boundary condition at each $\B_\v$:
\be
\label{IBC1}
(\dot{\mathcal{B}}_\v)^a e^i_a = 0.
\ee
Notice this condition also sets to zero any variation of the tangential frame-field component, i.e. $(\dot{\mathcal{B}}_\v)^a \delta e^i_a = 0$.

As mentioned in the introduction, particle masses represent the deficit angles of conical singularities in $\Sigma$. Up to $\SU(2)$-gauge transformations, these deficit angles determine the value of the holonomy around a loop encircling the particle (and only that particle). In general, the holonomy defined on a path $\gamma$ is given by the path-ordered exponential of the connection:
\be
\label{holonomy1}
h_\gamma:= \Pexp \int_\gamma \bm{A}.
\ee
Geometrically, this describes how a vector is rotated under parallel transport along $\gamma$.

Calculating the holonomy around a loop requires the specification of a base-point $b$ where the path integral begins and ends. Taking the path to begin at a point $b \in \B_\v$ and circle counterclockwise around the boundary, we can express the holonomy as:
\be
h_{\B_\v,b} = \mathbb{1} \cos \frac{m_\v}{2} - 2 \bm{u}_{\v,b} \sin \frac{m_\v}{2},
\ee
where $\bm{u}_{\v,b} \in \su(2)$ is a unit vector defining the axis of rotation (which depends upon the choice of base point $b$). See appendix \ref{appendix} for details in obtaining this form of the holonomy. 

To extract the gauge invariant mass $m_\v$, we take the trace of this equation to obtain the Wilson loop around the particle $\v$:
\be
\label{traceh}
W_\v = \Tr \ h_{\B_\v,b} = 2 \cos \frac{m_\v}{2},
\ee
where the dependence on the base-point has been traced out, i.e. $W_\v$ is the same for any choice of base point.

We now use Wilson loops to introduce mass-shell constraints\footnote{We borrow this name from \cite{Matschull}, although our constraints are written slightly differently.} for each particle, written in terms of the trace of holonomies $h_{\B_\v, b}$:
\be
\label{mass shell}
\psi_\v := 2 \cos^{-1} \frac{W_\v}{2} - m_\v,
\ee
imposing that the connection $\bm{A}$ encodes the deficit angle $m_\v$ associated with the mass of each particle. This constraint is the covariant analog of the 3d Riemannian spacetime relation $|p_\v|^2 = |m_\v|^2$ between 3d momentum and rest mass. Through these constraints, each particle adds a point of curvature to the spatial hypersurface giving it the geometry of a polyhedron. This implies that the total mass of all particles must be equal to $4\pi$ \cite{DJH}. Note that the upper limit for the deficit angle at a polyhedron vertex provides an upper limit on each particle mass $m_\v < 2 \pi$, which leaves no ambiguity in the inverse cosine used to define the mass shell constraints.

We can now incorporate point particles into the Hamiltonian by adding the mass-shell constraints to the pure gravity Hamiltonian:
\be
\label{H0}
H = \sum_\v \alpha_\v \psi_\v + \mathcal{F}(\bm{N}) + \mathcal{G}(\bm{\lambda}),
\ee
where $\alpha_\v$ is the Lagrange multiplier associated to the mass shell constraint $\psi_\v$.
This is again a sum of constraints as is the case for pure gravity. The phase space is given by $(\bm{A}, \bm{e}) \in \mathcal{P}$, but now these fields are defined on a two-sphere with punctures $\Sigma \cong S^2 \setminus \{\v \}$. The mass shell constraints impose conditions on the fields, imparting them with topological information associated to the particles. Note that the mass shell constraints are not first class with the Gauss constraint, so we shall need to do some additional work to obtain a first class constraint algebra. We discuss this in detail in the next section.

\subsection{Boundary terms and conditions}
\label{BTsec}
In calculating the various Poisson brackets and variations of the fields in order to find a consistent Hamiltonian theory, one generally finds boundary terms which must be properly dealt with to avoid an ill-defined variational principal and/or a second class constraint algebra. There are two ways in which such anomalous boundary terms can be dealt with: 1) add cancellation terms to the Hamiltonian; 2) place boundary conditions on the field variables and Lagrange multipliers which set the anomalous terms to zero. There is generally some freedom in this process; we will make a particular choice and stick with it for the remainder of the article. We point the interested reader to similar calculations done for the case of a bounded region of 4d Lorentzian gravity in terms of the Ashtekar variables \cite{HM}.

\subsubsection{Preserving the variational principal}
We begin by looking at the boundary terms and conditions necessary to preserve the variational principal. Variation of the Hamiltonian (\ref{H0}) results in a boundary term associated to each particle:
\be
\left. \delta H \right|_{\B_\v} =  \oint_{\B_\v} \left( \frac{\alpha_\v}{\sin(\frac{m_\v}{2})} \Tr \left( h_{\B_\v, x}\bm{\tau}^i \right) - N(x)^i \right) \delta A(x)^i - \oint_{\B_\v} \lambda (x)^i \delta e(x)^i,
\ee
where $h_{\B_\v, x}$ is the holonomy around the loop $\B_\v$ with base-point\footnote{Note that variation of a holonomy along a path $\gamma$ splits the path into two, i.e. $\delta h_\gamma = h_{\gamma_1} \delta \A h_{\gamma_2}$ where $\gamma_1$ is the part of the path before the variation, and $\gamma_2$ is the part of the path after the variation.} at the point of integration $x$. The second term vanishes due to the point particle condition (\ref{IBC1}). In order to eliminate the first term, we impose a condition at each particle boundary:
\be
N(x)^i =  \frac{\alpha_\v}{\sin(\frac{m_\v}{2})} \Tr \left( h_{\B_\v, x}\bm{\tau}^i \right) , \hspace{0.5in} \forall x \in \B_\v.
\ee
Using the equation (\ref{genh2}) we can write this as:
\be
\label{BCN1}
\N(x) = \alpha_\v \u_{\v, x}, \hspace{0.5in} \forall x \in \B_\v,
\ee
where $\u_x$ is the axis of rotation for $h_{\B_\v, x}$ (the holonomy around $\B_\v$ starting at the point $x$).
As we move around the boundary $\B_\v$, the vector $\N(x)$ will point in different directions although its magnitude $N$ remains constant.
Recall that $2 N^i:=e_0^i$ is the time component of the frame-field so that the choice of Lagrange multiplier $\alpha_\v$ has implications on the evolution of hypersurfaces. We shall make an appropriate choice once we have derived the equations of motion.

\subsubsection{Obtaining a first class constraint algebra}
Next we study the constraint algebra to determine which boundary conditions and/or additional boundary terms are required so that the constraints are first class.
Since the mass shell constraints and the flatness constraint do not contain the frame-field, the following Poisson brackets vanish trivially without need for additional boundary conditions:
\be
\left\{\psi_\v, \psi_{\v} \right\} = \left\{ \psi_\v, \mathcal{F}(\bm{N}) \right\} = \left\{ \mathcal{F}(\bm{N}), \mathcal{F}(\tilde{\bm{N}}) \right\} =0.
\ee

As mentioned above, the Gauss constraint has a non-trivial Poisson bracket with the mass-shell constraints:
\be
\left\{\mathcal{G}(\lambda), \psi_\v \right\} = \oint_{\B_\v} \left( \bm{u}_{\v,x} \right)^i (\rd_A \L)^i,
\ee
where $\bm{u}_{\v,x} \in \su(2)$ is a unit vector which points in the direction of particle momentum as seen from the point of integration $x$.
Details of this calculation are given in appendix \ref{aB}. In order to obtain a first class constraint algebra, we set the right hand side of the above equation to zero by imposing a condition at each particle boundary:
\be
\label{IBC3}
(\dot{\B}_\v)^a (\partial_a \lambda^i + \epsilon^{ijk}A^j_a\lambda^k) = 0,
\ee
where $(\dot{\B}_\v)^a$ is the vector tangent to the boundary. This condition imposes that the covariant external derivative of the Lagrange multiplier $\bm{\lambda}$, in a direction tangent to the particle boundary, vanishes at each boundary. In other words, $\bm{\lambda}$ is covariantly constant around each boundary $\B_\v$.

There are still two more Poisson brackets to check. Using the condition (\ref{IBC3}) along with the flatness and Gauss constraints, we find as desired that:
\be
\left\{ \mathcal{F}(\bm{N}), \mathcal{G}(\bm{\lambda}) \right\} = \left\{ \mathcal{G}(\bm{\lambda}), \mathcal{G}(\tilde{\bm{\lambda}}) \right\} = 0.
\ee

We can now summarize the results of our analysis. The full Hamiltonian with boundary terms is parameterized by the fields $(\bm{A},\bm{e})\in \mathcal{P}$ defined upon a sphere with punctures $\Sigma = S^2 \setminus \{ \v\}$, and is given by:
\be
H &=& \sum_\v \alpha_\v \psi_\v + \mathcal{F}(\bm{N}) + \mathcal{G}(\bm{\lambda}),
\ee
where we restate the constraints for convenience:
\be
\psi_\v &=& 2 \cos^{-1} \frac{W_\v}{2} - m_\v , \hspace{0.25in} \mbox{where} \hspace{0.25in} W_\v := \Tr \left( \Pexp \oint_{B_\v} \bm{A} \right), \\
\mathcal{F}(\bm{N}) &=& \int_\Sigma N^i \left( \rd A^i + \frac{1}{2} \epsilon^{ijk}A^j \wedge A^k \right), \\
\mathcal{G}(\bm{\lambda}) &=& \int_\Sigma \lambda^i \left( \rd e^i + \epsilon^{ijk} A^j \wedge e^k \right).
\ee
The field variables $(\bm{A},\bm{e})$ and Lagrange multipliers $\bm{N}$ and $\bm{\lambda}$ are subject to conditions at each particle boundary:
\be
&\N(x) = \alpha_\v \u_{\v, x} & \\[3pt]
&\dot{\B}_\v(x)^a \left( \partial_a \bm{\lambda} (x) + \left[ A(x)_a , \bm{\lambda}(x) \right] \right) = 0,& \\[3pt]
&\dot{\B}_\v(x)^a \e(x)_a = 0,&
\ee
for all points around the boundary $x \in \B_\v$.
With these conditions the variational principal is well defined and the constraint algebra is first class. Having established a consistent Hamiltonian, we move on to the next task: fixing a gauge for the field variables.

\section{Gauge fixing}

In this section we choose a gauge for the field variables, i.e. we specify a pair of fields $(\A, \e)$ such that $\bm{F}(x) = \bm{G}(x) = 0$ for all $x \in \Sigma$. In doing so, we will eliminate the kinematical constraints from the Hamiltonian according to Dirac's procedure \cite{Dirac}, leaving only the mass-shell constraints. These remaining constraints are the generators of dynamics. The gauge is fixed by defining the space manifold $\Sigma$ as a triangulation, so that we can solve the constraints within each triangle and allow for points of curvature (i.e. particles) at the vertices. This will allow us to give a piecewise definition of the field variables which can be glued together in a continuous manner.

Before we can specify the fields, there are some definitions we need to make clear. First, we give a precise definition of a CW complex, the topological space which underlies our field variables $(\A, \e)$. In two dimensions, a CW complex is a way of gluing together two-dimensional cells to form a closed topological space. This topology can support field variables which contain curvature singularities at a finite set of points. After defining this pre-geometry, we introduce a chart for each two-cell which permits us to write solutions for the fields $(\A, \e)$, cell-by-cell in a coordinate basis. Once these preliminary definitions are established, we then make a specific choice of fields and reduce the Hamiltonian according to the gauge-fixing procedure.

\subsection{CW Complex}

The spatial hypersurfaces that we shall work with are built upon CW complexes\footnote{The `C' is for closure-finite and the `W' is for weak topology.} \cite{Book2}. These are a rather general way of gluing together cells to form an $n$-dimensional topological spaces. A CW complex $C_n$ of dimension $n$
can be decomposed in terms of its $i$-skeletons ${C}_{i}$, $i=0,\cdots , n$.
Each ${C}_{i}$ is built from the ground up, defined recursively by gluing a disjoint union of $i$-dimensional open balls $\mathring{B}_{i}$, to ${C}_{i-1}$, where $C_0$ is a disjoint set of points.

Let us now consider in detail the two-dimensional CW complex and introduce a notation for the various objects. We begin with a set of points on a two-sphere $\{ \v \} \in S^2$. These points comprise the zero-skeleton $\Gamma^* \equiv \{ \v \}$ of the CW complex $\Delta$. We introduce {\it gluing maps} $s_\ed$ and define the one-skeleton $C_1$ by gluing one-dimensional open balls, called edges $\ed \equiv \mathring{B}_{1}$, to $\Gamma^*$:
\be
s_\ed: \partial \ed \to \Gamma^*.
\ee
These maps glue a set of edges $\{ \ed \}$ to the vertices $\v \in \Gamma^*$ by attaching the endpoints of the edges to the vertices in the zero skeleton. We require the gluing to be such that each edge is at attached to $\Gamma^*$ with different vertices at each end, each vertex is attached to an edge, and each vertex corresponds to the endpoint of at least three edges. The one-skeleton is defined as the union of the edges and vertices:
\be
C_{1}\equiv \left( \Gamma^* \underset{\ed}{\sqcup} \ed \right)/ \sim,
\ee
where the quotient by $\sim$ denotes the identification provided by the gluing maps: given $x\in \partial \ed$, $\v \in \Gamma^*$ we say that $ x\sim \v$ if
$s_\ed(x)=\v$.

At this point we have the one-skeleton $C_1$, a set of edges $\ed$ glued together at their endpoints $\v$. Moving up to the next dimension, we introduce gluing maps $s_c$ which glue two-dimensional open balls, called cells $c \equiv \mathring{B}_{2}$, to the one-skeleton $C_1$:
\be
s_c: \partial c \to C_1.
\ee
These maps glue a set of cells $c$ to the edges $\ed \in C_1$ such that the boundary of each cell $\partial c$ is attached to a set of closed edges $\bar{\ed} \equiv \ed \cup \partial \ed$ which form a closed loop. Each edge is in the boundary of two cells. The two-skeleton is defined as the union of cells, edges and vertices:
\be
\Delta \equiv \left( C_1 \underset{c}{\sqcup} c \right)/ \sim,
\ee
where the quotient by $\sim$ denotes here the following identification: given $x\in \partial c$, $y \in \bar{\ed}$ we say that $x \sim y$ if
$s_c(x)=y$.

The gluing maps $s_\ed$ and $s_c$ induce a map $s_{cc'}$ on each edge:
\be
\begin{array}{rl}
s_{cc'}:
 & \mathring{\ed}_c \to \mathring{\ed}_{c'} \\
 & \partial \ed_c \to \partial \ed_{c'}
\end{array}
\ee
Under this map, the edge $\ed_c \in \partial_c$ is identified with $\ed_{c'} \in \partial c'$ such that the interior of one edge is mapped to the interior of the other, and the endpoints of one edge are mapped to the endpoints of the other. 

The above definition is the general case of a two-dimensional CW complex. In order to provide the topology suitable for point particles on a sphere, we require that each edge $\ed_{cc'}$ is glued to a unique pair of cells $c$, $c'$. We furthermore take each cell $c$ to contain three edges $\ed$ in its boundary, so that we may conveniently picture each cell as a (topological) triangle.

Now, the spatial hypersurfaces $\Sigma$ we are dealing with are each homeomorphic to the punctured sphere $\Sigma_t \cong S^2 \setminus \{ \v\}$. We adapt the CW complex $\Delta$ to fit this by choosing the set of particles as the zero-skeleton $\{ \v \} \equiv \Gamma^*$. Given this zero-skeleton, there is some freedom in choosing a set of edges to connect these points. However, for a given number of particles $|\v|$, the Euler characteristic for polyhedra fixes the number of cells to be $|c| = 2(|\v|-2)$. Since the number of cells is fixed, different CW complexes stemming from different choices of edges are related by two-to-two Pachner moves\footnote{A two-to-two Pachner move, also called a bistellar flip, changes two cells into a new pair with a different adjacency relationship.}. An example of the CW complex $\Delta$ we use is shown in fig. \ref{justtriangles}.
\bcf
\includegraphics[width=0.4\linewidth]{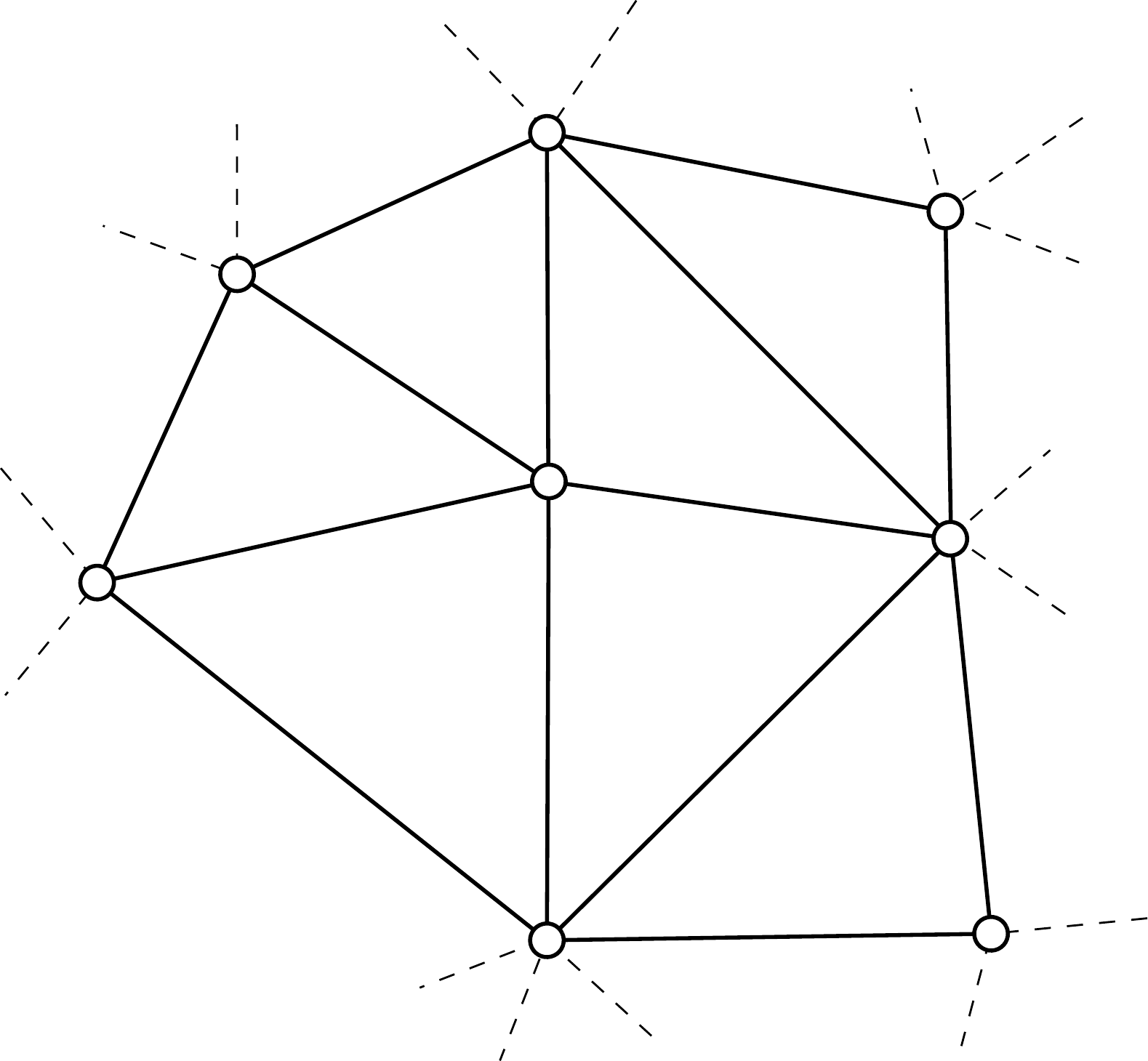}
\caption{\label{justtriangles}A neighborhood within the type of CW complex $\Delta$ used in this article, showing edges and vertices (where particles will reside). The dashed lines indicate that $\Delta$ continues outside of the neighbourhood illustrated here.}
\ecf

\subsection{Cell chart}

In order to define a chart, we begin by viewing the cell as a triangle defined in terms of a cartesian coordinate system $(x,y)$ for $\mathbb{R}^2$. There are three vertices $\v \equiv (x_\v, y_\v)$ with straight edges running between them, and a centroid at a point $\n \equiv (x_\n, y_\n)$. We shall use these coordinates when they are convenient, however they are not suitable for defining the smeared particle boundaries.

\begin{figure}[tb]
\begin{center}
\includegraphics[width=0.5\linewidth]{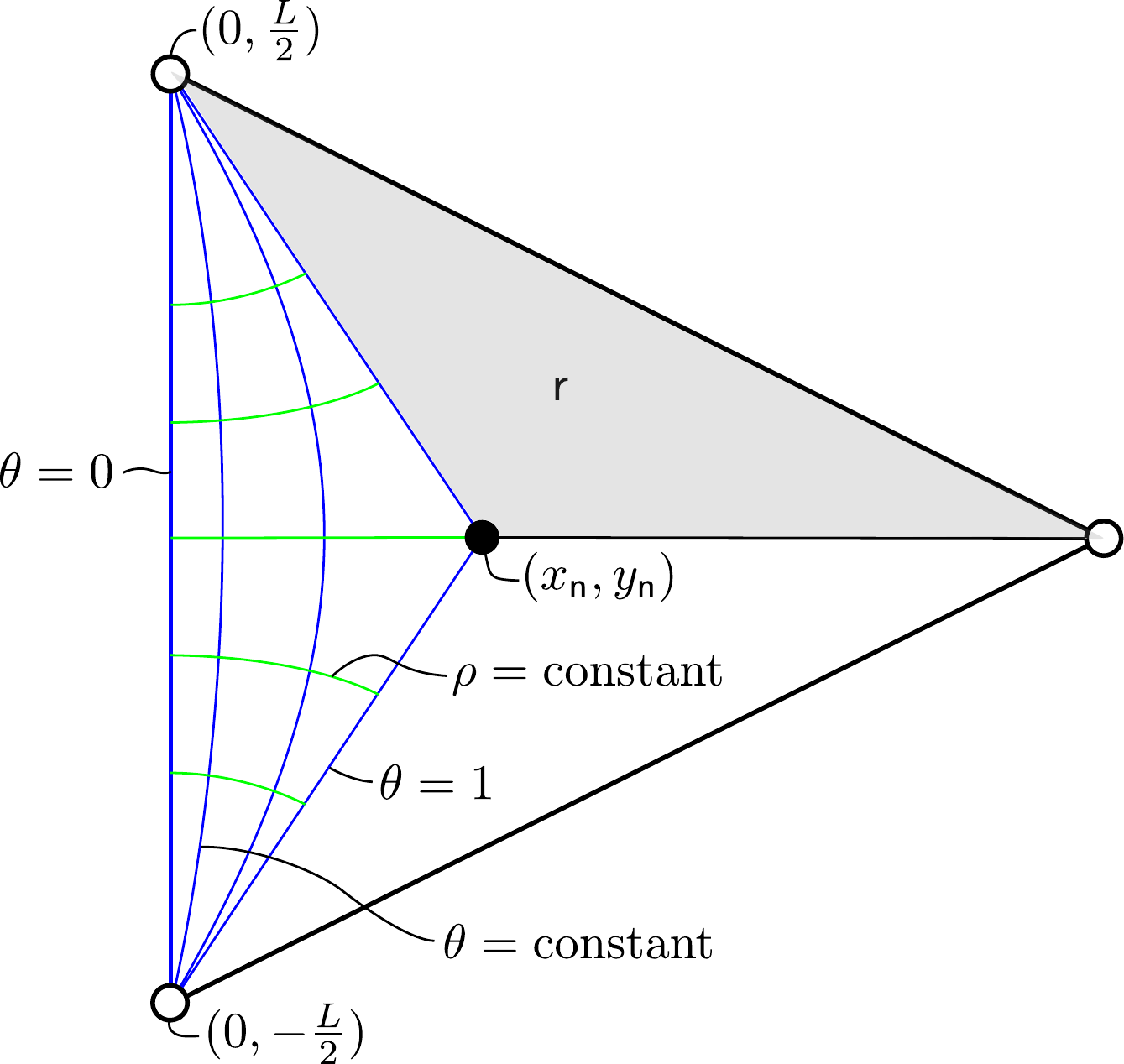}
\caption[Regions of a triangle]{A single triangle split into three regions by lines from the vertices (open circles) to the centroid (filled black circle). The $(\theta, \rho)$ coordinates are shown within one of the regions, with lines of constant $\theta$ shown in blue, and lines of constant $\rho$ in green. The line $\theta=0$ runs between the two vertices at $(0, \pm \frac{L}{2})$, and the line $\theta=1$ connects the centroid at $(x_\n, y_\n)$ to both of these vertices. A different region is labeled $\r$ and shaded in grey.}
\label{regions}
\end{center}
\end{figure}

We shall now introduce a new set of coordinates which allow us parameterize paths around particle boundaries, and will help us later on to write down an explicit form of the field variables $(\A, \e)$ which satisfy the constraints. The first step is to subdivide each triangular cell into three \textit{regions} $\r$ defined by edges joining the centroid $\n$ to the vertices $\v$ as shown in fig. \ref{regions}. The new coordinates $(\theta, \rho)$ provide an orthogonal basis for any single region and will allow us to define fields $(\A, \e)$ which are smooth over the entire cell.

Consider a single region $\r$ that has $\v$, $\v^\prime$, and the centroid $\n$ in its boundary. We assign a pair of cartesian coordinates $(x, y)$ to the region and place the origin at the midpoint of the edge $\ed_{\v \v'}$ between vertices $\v$ and $\v'$. In these coordinates, the centroid is at some $\n =(x_\n, y_\n)$, and the vertices are at $(0, \pm \frac{L}{2})$ where $L:= 2 y_\v = -2 y_{\v^\prime}$ is the coordinate length of $\ed_{\v \v'}$.

The first of our new coordinates is defined as follows:
\be
\theta(x,y) = \tanh \left( \tanh^{-1} (\theta^+) + \tanh^{-1} (\theta^-) \right), \qquad \qquad \theta^{\pm} = \frac{x}{x_\n} \pm 2 \left( \frac{y}{L}-\frac{x y_\n}{x_\n L} \right).
\ee
Over the range of values $\theta = [0,1]$, lines of constant $\theta$ provide a family of curves connecting the endpoints of the edge.
The line of $\theta=0$ is the edge $\ed_{\v \v'}$, and the line $\theta=1$ connects the vertices with the centroid. See fig. (\ref{regions}) for an illustration.

For the second coordinate $\rho$, we look for a function which together with $\theta$ provides an orthogonal basis. This means that the gradient of $\rho$ must be orthogonal to the gradient of $\theta$, i.e. $\rho$ must satisfy $\partial_x \theta \partial_x \rho + \partial_y \theta \partial_y \rho = 0$. Using the above definition of $\theta$, this equation boils down to:
\be
\label{rho_eq}
U \partial_x \rho + V \partial_y \rho = 0,
\ee
where we have defined:
\be
U:= x^2\left( y_\n^2-\left(\frac{L}{2}\right)^2 \right)-x_\n^2 \left( y^2 - \left(\frac{L}{2}\right)^2 \right), \qquad \qquad V:= 2 x x_\n (x_\n y - y_\n x).
\ee

In general the solution to (\ref{rho_eq}) must be found numerically. However, when the centroid is positioned on the $x$-axis so that $y_\n=0$, we can find an analytical solution:
\be
\rho = C_1 y^{\frac{L^2}{4x_\n^2}} \left( (x-x_\n) (x+x_\n) + \frac{4 x_\n^2 y^2}{8 x_\n^2+L^2} \right) + C_2,
\ee
where $C_1, C_2$ are arbitrary constants. In any case, we shall denote as $\rho_\v$ the value of $\rho$ along the particle boundary where $\B_\v \bigcap \r$. 

We now have a set of coordinates $(\theta_\r, \rho_\r)$ for each of the three regions $\r$ in a cell. Recall that particle boundaries are smeared in a polar coordinate system $(r,\phi)$ by not identifying points with different $\phi$-values in the limit $r \rightarrow 0$, except for the identification $\phi = \phi + 2\pi$. In terms of $(\rho, \theta)$ coordinates, the smearing implies that we do not identify points with different $\theta$-values when $\rho=\rho_\v$. Within a region, one can define a path which partially goes around a particle boundary by following $\theta$ over the range $[0,1]$ while keeping fixed $\rho=\rho_\v$. Considering all of the regions which intersect at a particle boundary, one can define a set of such paths which join end to end and go completely around the particle boundary. This is useful for defining the fields $(\A,\e)$ at the end of this section, and for reducing the Hamiltonian in the next section.

\subsection{Solutions to the constraints}

Now that we have a definition of the CW complex $\Delta$ and a useful set of coordinate patches for each cell, we can go ahead and make a specific choice of fields which satisfy the constraints. Within each cell $c$, we must find $(\A_c(x),\e_c(x))$ such that $\bm{F}(x) = \bm{G}(x) = 0$ for all $x \in c$. The general solution is given by a pair of functions, a `rotation' function $a_c(x) \in \SU(2)$ and `coordinate' function ${\bm z}_c (x) \in \su(2)$:
\be
\label{solutions}
\A_c = a_c \rd a_{c}^{-1}, \hspace{1in} \e_c = a_c \left( \rd {\bm z}_c \right) a_{c}^{-1}.
\ee

We use a piecewise definition to give the fields over all of $\Sigma \equiv \Delta \setminus \{ \v \}$: $\A \equiv \cup_c \A_c$ and $\e \equiv \cup_c \e_c$ for all $c \in \Delta$. The continuity of these piecewise fields must be defined by a limiting procedure, where one ensures that the value of $(\A_c, \e_c)$ agrees with the value of $(\A_{c'}, \e_{c'})$ as one approaches the edge $\ed_{cc'}$ from either side. The connection and frame-field are continuous throughout $\Sigma$ so long as there exists a constant gluing element $h_{c c'} \in \SU(2)$ associated to each edge such that for the coordinate functions:
\begin{subequations}
\label{gluing_z}
\be
\lim_{x' \to \ed_{c'c}} {\bm z}_{c'}(x') &=& \lim_{x \to \ed_{cc'}} h_{c c'}^{-1}  \left({\bm z}_{c}(x) - \bm{b}_{c c'} \right) h_{c c'} , \\
\lim_{x' \to \ed_{c'c}} \rd {\bm z}_{c'}(x') &=& \lim_{x \to \ed_{cc'}} h_{c c'}^{-1} \rd {\bm z}_{c}(x) h_{c c'} ,
\ee
\end{subequations}
where the constant $\bm{b}_{c c'} \in \su(2)$ is a translation\footnote{The translation $\bm{b}_{c c'}$ is the covariant integral along a path $\gamma$ from the origin of $\z_c$ to the origin of $\z_{c'}$, i.e. $\bm{b}_{c c'} = \int_\gamma h_\pi^{-1} \e h_\pi$ where $h_\pi$ is the holonomy along a set of paths $\pi$ from the origin of $\z_c$ to the points of integration. Since the connection is flat, any paths $\gamma$ and $\pi$ within the union $c \bigcap c'$ will yield the same result.}. In the above notation we have that $x \in c$ and $x' \in c'$, while the edge $\ed_{cc'}$ seen from cell $c$ is identified with the edge $\ed_{c'c}$ seen from cell $c'$. The first equation shows that along the edge $\ed_{cc'}$, the coordinates $\z_c$ are related to $\z_{c'}$ by a Poincar\'e transformation.
Now, for the rotation functions we must have:
\begin{subequations}
\label{gluing_a}
\be
\lim_{x' \to \ed_{c'c}} a_{c'}(x') &=& \lim_{x \to \ed_{cc'}} a_{c}(x) h_{c c'} , \\
\lim_{x' \to \ed_{c'c}} \rd a_{c'}(x') &=& \lim_{x \to \ed_{cc'}} \rd a_{c}(x) h_{c c'} . 
\ee
\end{subequations}
Notice the first equation gives an expression for the gluing elements in terms of the rotation fields evaluated at the edge:
\be
\label{h(a)}
h_{c c'} = a_{c}(x)^{-1} a_{c'}(x),
\ee
which must be constant for all $x \in \ed_{cc'}$.

\subsubsection{Coordinate functions $\z_c$}

Let us now define the coordinate functions more precisely. Recall that on each particle boundary $\B_\v$, the component of the frame-field tangent to $\B_\v$ must vanish (\ref{IBC1}). This translates into a boundary condition on ${\bm z}_c$:
\be
\label{IBC1b}
\dot{\B}_\v(x)^a \partial_a \bm{z}_c(x) = 0 \hspace{0.5in} \forall x \in \B_\v \bigcap c.
\ee
We require each function $\bm{z}_c$ to be single valued in $c$ so that the metric $g_{ab} = \partial_a \z \cdot \partial_b \z$ is non-degenerate, except at the particle boundaries where this condition is telling us that $\bm{z}_c$ must be constant on the boundary $\B_\v$. This means that the functions $\bm{z}_c$ can serve as coordinates within each cell since each point receives a unique value. Degeneracy at $\B_\v$ means that the vertex $\v$ receives a single coordinate value $\z_c(\v)$ within a cell even though the boundary is smeared\footnote{Writing $\z_c(\v)$ is an abuse of notation since the point $\v$ has been excised and the function $\z_c$ is not defined at this point. By this notation we mean any point along the intersection $\B_\v \bigcap c$. The choice is arbitrary due to the condition (\ref{IBC1b}).}.

A suitable choice of coordinate function is given in terms of cartesian coordinates by:
\be
\label{z1}
\bm{z}_{c} = x_c \x_c + y_c \y_c,
\ee
where $\x_c$ and $\y_c$ are $\su(2)$ basis elements representing unit vectors in $\mathbb{R}^3$ associated to the $(x_c,y_c)$ coordinates. These unit vectors span a plane, and we denote the direction perpendicular to this plane as $\ti_c = [\x_c, \y_c]$. We can see that the coordinate function $\z_c$ is giving the (topological) cell $c$ the geometry of a triangle within $\mathbb{R}^3$. The gluing element $h_{cc'}$ rotates the plane spanned by $(\x_c, \y_c)$ into the plane spanned by $(\x_{c'}, \y_{c'})$, and the coordinate $\z_c$ is related to the neighbouring $\z_{c'}$ by a Poincar\'e transformation (a rotation by $h_{cc'}$ and a translation by $\b_{cc'}$). Each coordinate function then serves as a reference frame for the associated cell with the usual transformation between reference frames in flat space. Note that each vertex is shared by at least three cells and is given a different coordinate $\z_c(\v)$ within each of these cells. This is a direct result of the coordinate functions $\z_c$ acting as references frames, since different frames see the particle in different locations.

\subsubsection{Rotation functions $a_c$}

To specify the rotation functions we make use of the $(\theta_\r, \rho_\r)$ coordinates, defined separately in each of the three regions $\r$ of a cell $c$. We shall provide a definition of $a_c$ which joins smoothly at the intersection of regions within a cell (i.e. the $\theta = 1$ lines). In order to do this, we introduce a normalized `bump' function $f(\theta)$ which satisfies $\int_0^1 f(\theta) \rd \theta = 1$, and goes to zero smoothly as $\theta \rightarrow 0$ and as $\theta \rightarrow 1$. An example of a bump function is:
\be
f(\theta) =
\left\{
\begin{array}{cl}
C e^{\frac{1}{(2 \theta - 1)^2 - 1}}, & 0 \le \theta \le 1 \\
0, & \mbox{otherwise}
\end{array}
\right. ,
\ee
where $C^{-1} = \int_0^1 e^{\frac{1}{(2 \theta - 1)^2 - 1}} \rd \theta$.

Now, we define the rotation function within a region as:
\be
\label{gena}
a_{\r}(\theta) = \exp \left( -\bm{P}_{\r} \int_1^\theta f(\tilde{\theta}) \rd \tilde{\theta} \right).
\ee
When $\theta=0$, the $\SU(2)$-valued function $a_{\r}$ defines a rotation about a direction that is given by $-\bm{P}_{\r}/|P_{\r}|$, while the angle of rotation is $|P_{\r}|$. Because the bump function goes to zero smoothly as $\theta \rightarrow 0$, we have that $a_{\r}(\theta) \rightarrow e^{-\bm{P}_\r}$ and $\rd a_{\r}(\theta) \rightarrow 0$ smoothly as one approaches the edge associated to $\r$.

The rotation function over the cell is given piecewise from the definitions in each region:
\be
a_c = \bigcup_\r a_\r .
\ee
Along the intersection between two regions $\r \bigcap \r^\prime$ where $\theta = \theta^\prime = 1$, we have that $a_{\r}(\theta) = a_{\r^\prime}(\theta^\prime) = \mathbb{1}$. Due to the smoothness of the bump functions, the rotation function $a_c$ is smooth across these intersections, and throughout the entire cell.

In order that the gluing conditions (\ref{gluing_a}) are satisfied, the product of rotation functions must be constant $h_{c c'}$ along each edge. Consider two neighbouring triangles $c$, $c'$ and the shared edge $\ed_{c c'}$. Notice that since the rotation function defined in (\ref{gena}) is a function of $\theta$ only, it is constant along the edge of the triangle. This means that $a_c$ and $a_{c'}$ are each constant along the edge $\ed_{c c'}$, and so then is their product $a_c^{-1}(x) a_c(x) = h_{cc'}$ for all $x \in \ed_{cc'}$ as required.

\subsubsection{Comparison with the model of Deser, Jackiw and 't Hooft}

The seminal work on point particles in 3d gravity was given by Deser, Jackiw and 't Hooft (DJH) in \cite{DJH}. The DJH model is built upon Minkowski spacetime with wedge-shaped portions removed to account for the deficit angles associated to particle masses. Each particle wordline is on the edge of a wedge which extends to infinity, and the sides of each wedge are identified. With this identification, the slice of a wedge lying within a spatial hypersurface appears as a one-dimensional edge (which we will here call a tail) that spans a half-line to infinity. Identifying either side of the tail leads to matching conditions for the coordinates along the tail. It is instructive to reproduce the DJH matching conditions within our framework. We shall consider two of the cases considered in the article \cite{DJH}: 1) a single particle; 2) a pair where one particle moves relative to the other with the tails on top of each other.

In the present article we can take the coordinate functions $\z_c$ as local patches of flat space. The gluing conditions (\ref{gluing_z}) for the coordinate functions are then the analog of the DJH matching conditions for the flat space coordinates on either side of a tail. In order to create the scenarios relevant for comparison with the DJH matching conditions, we can consider certain CW complexes with specific choices of gluing elements $h_{cc'}$ associated to edges.

Let us first consider the case of a single particle at rest. The tail stemming from a particle in the DJH model looks in our case like a vertex $\v$ that has a non-trivial gluing element $h_{cc'}$ on only one edge $\ed_{cc'} \ni \v$, while the other edges which end at this vertex carry trivial gluing elements. At some point $x \in \ed_{cc'}$ which maps to the point $x' \in \ed_{c'c}$, we have the following condition on coordinate functions:
\be
\label{genz}
\z^{c'}(x') = h_{cc'}^{-1} \z^c(x) h_{cc'}.
\ee
Since the particle is at rest, the gluing element must describe a rotation about the direction $\ti_c = \ti_{c'}$ by an angle given by the mass:
\be
h_{cc'} = e^{m_\v \ti_c} .
\ee
These last two equations are the $\SU(2)$ equivalents of the equations numbered (5.1) in \cite{DJH}. 

\begin{figure}[tb]
\begin{center}
\includegraphics[width=0.5\linewidth]{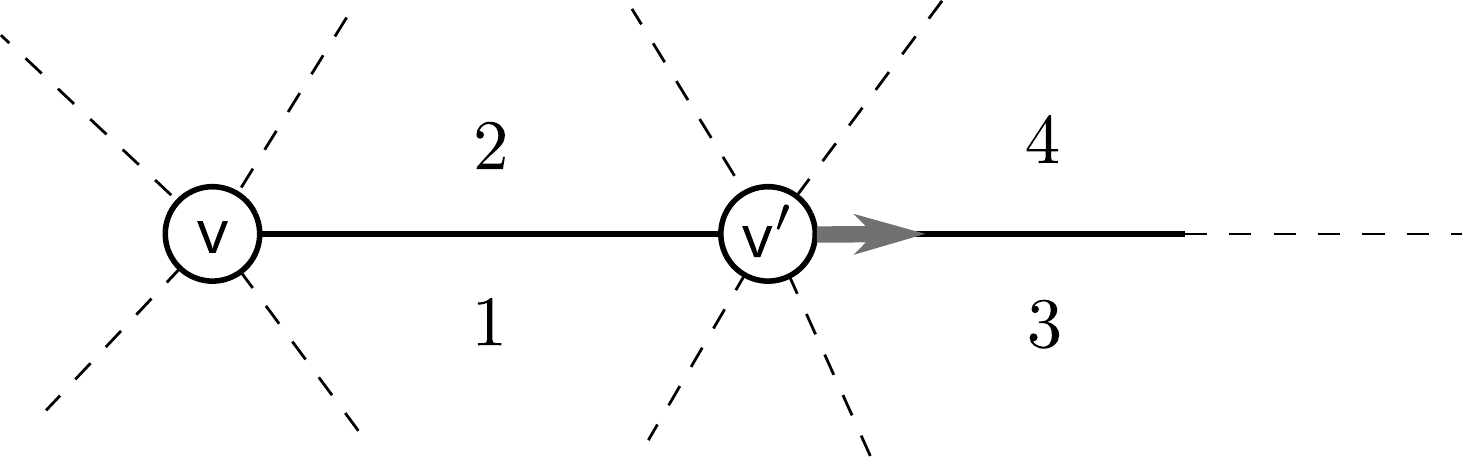}
\caption[Triangulation for DJH matching conditions]{Triangulation in the neighbourhood of two vertices $\v$ and $\v'$ with holonomies chosen appropriately for the matching conditions of \cite{DJH} to be applied. The particle $\v'$ moves to the right along the edge $\ed_{34}$.}
\label{DJH}
\end{center}
\end{figure}

Now let us consider the second scenario where one particle moves relative to another with one tail on top of the other. To set this up within our model we consider that the edge $\ed_{12}$ joins particles $\v$ and $\v'$ as in fig. \ref{DJH}, where the gluing element $h_{12} = e^{m_\v \ti_1}$ is a rotation and $h_{34} = e^{\p_{\v'}} e^{m_\v \ti_1}$ is a rotation and boost for some $\p_{\v'} \in \su(2)$. We take all other gluing elements on edges connected to these particles to be trivial. To mimic the coordinate labels used in \cite{DJH}, we choose origins for the coordinate functions such that $\z_1(\v) = \z_2(\v) = 0$ and $z_3(\v') = z_4 (\v') = 0$. We also allow the definition of $\z_1$ to extend into cell $3$, and we extend $\z_2$ to be defined within cell $4$. According to our gluing rules, these functions then satisfy $\z_4 = \z_2 - \b_{21}$ and $\z_3 = \z_1 - \b_{12}$, where the translations between reference frames are given by the coordinate length of $\ed_{12}$ as seen from either side of the edge, i.e. $\b_{12} = \z_1(\v') - \z_1(\v)$ and $\b_{21} = \z_2(\v') - \z_2(\v) = h_{12}^{-1} \b_{12} h_{12}$.

Let us now compare with the DJH model. At some point $x \in \ed_{34}$ which is mapped to $x' \in \ed_{43}$ we have the gluing relation:
\be
\z_4(x') &=& h_{34}^{-1} \z_3(x) h_{34} \nonumber \\
\z_2(x') - \b_{21} &=& h_{34}^{-1} (\z_1(x) - \b_{12}) h_{34} \nonumber \\
\z_2(x') &=& h_{12}^{-1} \b_{12} h_{12} + h_{34}^{-1} (\z_1(x) - \b_{12}) h_{34} \nonumber \\
\z_2(x') &=& e^{-m_\v \ti_1} (\b_{12}  +  e^{-\p_{\v'}} ( \z_1(x) - \b_{12} ) e^{\p_{\v'}}) e^{m_\v \ti_1}.
\ee
This is the $\SU(2)$ analog of equation (5.8) in \cite{DJH}.

We have shown that the coordinate functions $\z_c$ together with the gluing elements $h_{cc'}$ are able to describe the matching conditions given in the DJH model. Each $\z_c$ is a local patch of flat space that is related to neighbouring patches by translations and $\SU(2)$ rotations. This presents a picture of the geometry in terms of flat triangles that are glued together at their edges by rotations.

The geometry in terms of these flat triangles defined by the $\z_c$ coordinates is a discontinuous geometry, with the gluing elements representing a connection that is non-zero only on the edges. In general relativity we usually require continuous geometries. The Hamiltonian we are working with is given in terms of a connection $\A$ and frame-field $\e$, and using the constraint solutions (\ref{solutions}) we can give a continuous geometry in terms of $(\A, \e)$ for point particles in 2+1 dimensions. We have so far given the definitions for describing a gauge choice of fields $(\A, \e)$, but we have yet to implement this choice in the Hamiltonian system of equations. This is done in the next subsection.

\subsection{Hamiltonian gauge reduction}

We now employ the formalism of Dirac \cite{Dirac} to choose specific $(\A, \e)\in \mathcal{P}$ which satisfy the constraints. We have $6 \times 2$ degrees of freedom per point to be fixed in the variables $(A^i_a, e^i_a)$, while each of the constraints, $F^i$ and $G^i$, can be used to fix $3 \times 2$ degrees of freedom per point. We shall do this in two steps, first eliminating the Gauss constraint then the flatness constraint.

The first condition is:
\be
\label{gf1}
\bm{\mathcal{C}}_1:=\bm{A}_\theta - a \partial_\theta a^{-1}.
\ee
This constraint is applied within each region $\r \in \Sigma$.
A good gauge fixing condition must be second class with at least one of the constraints, and this condition is second class with the Gauss constraint. The Poisson bracket is:
\be
\left\{ (\mathcal{C}_1)^i, \int_\Sigma \lambda^{(j)} G^{(j)} \right\} = -\delta^{i(j)} \left(\partial_\theta \lambda^{(j)} + \epsilon^{(j)kl} A_\theta^k \lambda^l \right),
\ee
where we do not sum over the bracketed index $(j)$ so that the right hand side is a $3 \times 3$ diagonal matrix giving the Poisson brackets between each component of the gauge condition with each component of the Gauss constraint.

The condition must be preserved dynamically, which means the evolution equation provided by the Poisson bracket with the Hamiltonian $H_f$ must vanish. Calculating this we obtain:
\be
\label{CC1}
\left\{ \bm{\mathcal{C}}_1, H_f \right\} = \partial_\theta \bm{\lambda} + \left[ \bm{A}_\theta, \bm{\lambda}\right] = 0.
\ee
Setting the right hand side to zero provides a condition on $\bm{\lambda}$.

Following the gauge fixing procedure, we now define Dirac brackets. In general, the Dirac bracket for two functions $f$ and $g$ of the phase space variables is defined as:
\begin{equation}
\label{dbrack}
\left\{f,g \right\}_D := \left\{f,g\right\} - \left\{f,\Phi_m\right\} (M^{-1})^{mn}\left\{\Phi_n,g\right\} ,
\end{equation}
where $\Phi_m$ are the constraints, and the antisymmetric matrix $M$ is defined as:
\begin{equation}
M_{mn} := \left\{ \Phi_m, \Phi_n \right\}.
\end{equation}
Notice that the invertibility of this matrix depends upon the set of constraints $\Phi_m$ being second class with each other.

We can set the Gauss constraint and the gauge condition $\bm{\mathcal{C}}_1$ \textit{strongly} to zero, so long as we use Dirac brackets instead of Poisson brackets. In doing so, the gauge-fixed variable $\bm{A}_\theta$ and its complex conjugate $\bm{e}_\rho$ become non-dynamical, and we can eliminate them from the Hamiltonian since they are now fixed in terms of ($\bm{z}$, $a$) and the remaining field variables ($\A_\rho, \e_\theta$) through the Gauss constraint and the gauge condition. The matrix $M$ is easily inverted, and one can check that for the remaining phase space variables, the Dirac brackets (\ref{dbrack}) are equivalent to Poisson brackets.

We have now partially fixed our gauge and can continue using Poisson brackets in our analysis. The Gauss constraint has been eliminated and the partially fixed Hamiltonian is given by:
\be
H_{pf} = \sum_\v \alpha_\v \psi_\v + \mathcal{F}(\bm{N}) .
\ee
We fix the remaining degrees of freedom with the condition:
\be
\label{gf2}
\bm{\mathcal{C}}_2:=\bm{e}_\theta - a \left( \partial_\theta \bm{z} \right) a^{-1}.
\ee
This is second class with the flatness constraint:
\be
\left\{\left(\mathcal{C}_2 \right)^i_\theta, \int_\Sigma N^{(j)} F^{(j)} \right\} = \delta^{i(j)} \left(\partial_\theta N^{(j)} + \epsilon^{(j)kl} A_\theta^k N^l \right).
\ee
The same procedure as above involving Dirac brackets can be applied again, so that we can set $\left(\mathcal{C}_2\right)^i$ and $F^i$ strongly to zero and eliminate the remaining degrees of freedom in the field variables.

Preserving this constraint dynamically provides a condition on the Lagrange multiplier $\bm{N}$. Using the partially fixed Hamiltonian, the condition is:
\be
\label{CC2}
\left\{ \bm{\mathcal{C}}_2, H_{pf} \right\} = \partial_\theta \bm{N} + \left[ \bm{A}_\theta, \bm{N}\right] = 0.
\ee

The second gauge choice removes the flatness constraint and reduces the Hamiltonian to:
\be
\label{H_R}
H_R = \sum_\v \alpha_\v \psi_\v
\ee
The connection $\bm{A}$ and frame-field $\bm{e}$ are now completely determined by the constraints and the gauge choices (\ref{gf1}, \ref{gf2}). The topological degrees of freedom associated with the particles are contained within the parameters $\bm{P}_{\r}$ defining the $a$-fields in (\ref{gena}). Notice the Hamiltonian has support at the particles so we expect some non-trivial action at these locations.

The gauge fixing procedure places conditions on the Lagrange multipliers, and we now present a solution which satisfies these conditions.
We found above that $\N$ must satisfy (\ref{CC2}), while also satisfying at the particle boundaries:
\be
\N(x) =  \alpha_\v \u_{\v, x}  \hspace{1in} \forall x \in \B_\v.
\ee
A solution for the Lagrange multiplier within each region $\r$ is given by:
\be
\bm{N}_{\r}(x) = a_{\r}(\theta) \bar{\bm{N}}_{\r}(\rho) a_{\r}(\theta)^{-1}, \hspace{0.5in} \forall x \in \r,
\ee
where $\bar{\N}_\r(\rho)$ is a function of $\rho$ only.
Recall that the two particles $\v, \v'$ associated to a region sit at $\rho_\v$ and $\rho_{\v'}$ respectively, where $\rho_{\v'} < \rho_\v$. We define at each particle boundary:
\be
\label{BCN}
\bar{\bm{N}}_{\r}(\rho_\v) =  \alpha_\v \u_{\v, b}, \hspace{0.5in} \bar{\bm{N}}_{\r}(\rho_{\v'}) = -\alpha_{\v'} \u_{\v', b'},
\ee
where $b$ (resp. $b'$) is the intersection between the $\theta=1$ line and $\B_{\v}$ (resp. $\B_{\v'})$. To specify this function over the values of $\rho$ ranging between the particles we use a `double-bump' function, this time taking $\rho$ as the argument. Let us label $\rho = \rho_\n$ as the line of constant $\rho$ which intersects the centroid of the cell. The function $f(\rho)$ is normalized so that:
\be
\int_{\rho_\n}^{\rho_\v} f(\rho) \rd \rho = \int_{\rho_{\v'}}^{\rho_\n} f(\rho) \rd \rho = 1,
\ee
and goes smoothly to zero in the limits $\rho \rightarrow \rho_\v$, $\rho \rightarrow \rho_{\v'}$ and $\rho \rightarrow \rho_\n$. In other words, we have one normalized bump over the range $\rho_\v < \rho < \rho_\n$ and another between $\rho_\n < \rho < \rho_{\v'}$. We can now define $\bar{\bm{N}}_{\r}$ throughout the region:
\be
\bar{\bm{N}}_{\r}(\rho) = \ti + \left\{
\renewcommand{\arraystretch}{2.5}
\begin{array}{ll}
(\alpha_\v \u_{\v, b} - \ti) \displaystyle \int_{\rho_\n}^{\rho} f(\tilde{\rho}) \rd \tilde{\rho} , & \rho_\n \le \rho \le \rho_\v \\
(\alpha_{\v'} \u_{\v', b'} - \ti) \displaystyle \int_{\rho}^{\rho_\n} f(\tilde{\rho}) \rd \tilde{\rho} , & \rho_{\v'} \le \rho < \rho_\n
\end{array}
\right.
\ee
Values of this function vary smoothly from $\alpha_{\v'} \u_{\v', b'}$ to $\ti$ to $\alpha_\v \u_{\v, b}$ as one travels from $\B_{\v'}$ to the line $\rho = \rho_\n$ and then to $\B_\v$.
With this definition, the Lagrange multiplier in the region $\N_\r = a_\r \bar{\N}_\r a_\r^{-1}$ satisfies the boundary conditions at $\B_\v$ and $\B_{\v'}$. Moreover, one can check that this definition yields a smooth function over the triangle $\N_c = \cup_{\r \in c} \N_\r$, and also a smooth function over all of $\Sigma$ with the piecewise definition $\bm{N} = \cup_{c \in \Sigma} \N_c$. As for $\alpha_\v$, we shall fix this below once we have the equations of motion.


The other Lagrange multiplier $\bm{\lambda}$ must satisfy the condition (\ref{CC1}) for all $x \in \Sigma$. Any function of the form $a \bar{\L} a^{-1}$ for a function of $\rho$ only $\bar{\L}(\rho) \in \su(2)$ will do the trick, but for concreteness we choose:
\be
\bm{\lambda} = \bigcup_{c \in \Sigma} \bigcup_{\r \in c} a_\r \bm{\tau}^0 a_\r^{-1}.
\ee

We now summarize what we have accomplished with this gauge fixing procedure. We have a spatial manifold defined as a two-dimensional CW complex with the vertices removed $\Sigma = \Delta \setminus \{\v\}$. We have a particle at each vertex $\v \in \Gamma^*$ of the zero-skeleton. The frame-field and connection have been specified (gauge-fixed) within each cell in terms of coordinate and rotation functions $(a_c, \z_c)$ according to (\ref{solutions}). These definitions depend upon a choice of bump function $f_\r$ and rotation parameter $\bP_\r$ within each region of each cell. The solutions are glued together along each edge $\ed$ in the one-skeleton by a constant $h_\ed \in \SU(2)$ and the rules (\ref{gluing_z}, \ref{gluing_a}) to provide continuous fields $(\A, \e)$ over all of $\Sigma$. The input to this procedure is the choice of CW complex $\Delta$, bump functions $f_\r$ and rotation parameters $\bP_\r$.

Fixing the gauge amounts to choosing a specific point $(\A, \e)$ in the constrained subspace:
\be
\label{C}
\mathcal{C}^G_{\Gamma^*} = \left\{ (\A, \e)\in \mathcal{P} \ | \ \mathcal{F}[\bm{N}] = \mathcal{G}[\bm{\lambda}] = 0 \right\},
\ee
where the subscript $\Gamma^*$ indicates the set of points which have been excised.
This constrained subspace contains all of the physical degrees of freedom associated to the particles. This phase space contains points which are related by gauge transformations (those generated by the flatness and Gauss constraints). If we identify all points related by gauge transformations, we obtain the reduced phase space:
\be
\label{P}
\mathcal{P}^G_{\Gamma^*}  = C_{\Gamma^*} / (\mathcal{F} \times \mathcal{G}) = \mathcal{P} \sslash (\mathcal{F} \times \mathcal{G}),
\ee
where we have used a double-slash notation on the right hand side: the first slash is for applying the constraints, and the second slash is for `modding out' the gauge transformations. Any pair of fields $(\A, \e)$ which satisfy the constraints (i.e. a point in $\mathcal{C}_{\Gamma^*}$) can act as a representative of the equivalence class $[(\A, \e)] \in \mathcal{P}_{\Gamma^*}$ related by gauge transformations.

\section{Particle degrees of freedom}

Now that we have fixed a gauge and reduced the Hamiltonian, how can we extract information about the particles from the fields $(\A, \e)$? In other words, what is the precise way in which the particle positions and velocities are seen in the gravitational field?

Let us first consider how to define the particle positions. In a relativistic theory, one cannot say anything about position without specifying a frame of reference. We are working within a CW complex, where each cell $c$ possesses its own coordinate function $\z_c$ giving each cell the geometry of a triangle. We can use the centroid $\n_c$ of a cell as the point of reference for defining the location of the vertices $\v \in \partial c$. The position of the particle $\v$ seen in the frame of $c$ is:
\be
\label{q2}
\q^c_\v := \int_{\n_c}^{\v} h_\pi^{-1} \e h_\pi = \int_{\n_c}^{\v} \rd \z = \z(\v) - \z(\n_c),
\ee
where $\pi$ is a path from the centroid to the vertex, and $h_\pi = a^{-1}_c$. Note that since the connection is flat, the holonomy $h_\pi$ is the same for any path from the centroid to the vertex. Using the gluing rules, one finds for neighbouring cells $c$ and $c'$ that:
\be
\label{qrel}
\q^{c'}_\v = h_{cc'}^{-1} \left( \q^c_{\v} - \b_{cc'} \right) h_{cc'}.
\ee

Next we turn to the momentum. We want something defined in terms of the connection so that it has a non-trivial Poisson bracket with $\q_\v^c$. The obvious choice is the holonomy around the particle:
\be
\label{p1}
\p^c_\v := \Pexp \int_{\B_\v, b} \A = h_{cc'} h_{c'c''} \cdots h_{c''' c} ,
\ee
where on the right hand side we have a counter-clockwise ordered product of the gluing elements, beginning and ending at cell $c$. In this definition we have chosen the base-point $b$ of the loop $\B_\v$ to be at the intersection of $\B_\v$ and the $\theta=1$ line. Momenta of neighbouring cells are related by:
\be
\label{prel}
\p^{c'}_\v = h_{cc'}^{-1} \p^c_{\v} h_{cc'}.
\ee
Only the orientation of a momentum is affected by such a rotation, so that the particle masses are the same as seen from any cell:
\be
m_\v = 2 \cos^{-1} \frac{\Tr (\p^c_\v)}{2} ,
\ee
as they should be.

The Poisson algebra of these relative position and momentum variables is given by:
\be
\label{qpPB}
\{ \q^c_\v, \q^{c'}_{\v'} \} = [\q^c_\v, \q^{c'}_{\v'}] \delta_{\v \v'} , \qquad  \{ \p^c_\v, \p^{c'}_{\v'} \} = 0, \qquad \{ (q^c_\v)^i, \p^{c'}_{\v'} \} = -\t^i \p^c_\v \delta_{\v \v'} ,
\ee
where we used (\ref{prel}) to calculate the last bracket\footnote{In defining the last bracket, we also need a precise definition of the Poisson bracket $\{\e(x), h_{\B_\v, b}\}$ when $x$ is at the base-point (both the start and end of the loop). Here we define this intersection to be at the start of the loop. See equation (\ref{ehPB}) and the footnote which follows this equation for details.} and the first bracket follows from the Jacobi identity.

The variables $(\q_\v^c, \p_\v^c)$ 
provide a position and momentum for each particle $\v$ as seen from each cell $c$ which contains $\v$. The reduced Hamiltonian can be written in terms of the momenta only:
\be
\label{parH}
H = \sum_\v \alpha_\v \psi_\v, \qquad \psi_\v = 2 \cos^{-1} \frac{\Tr (\p_\v)}{2} - m_\v ,
\ee
where we have not labeled the momenta $\p_\v$ with a cell $c$, since this choice is irrelevant under the trace.
This Hamiltonian has support on each particle boundary, and we anticipate that it will generate dynamics at the vertices of each triangle.

\section{Dynamics on the CW complex}

Now that we have established the kinematics and arrived at a reduced Hamiltonian, let us turn to the dynamics. The first step is to define initial data. From the previous discussion it is clear that we cannot simply assign a position and momentum to each particle. We will need to have a pair $(\q_\v^c, \p_\v^c)$ for each cell $c$ which contains the vertex $\v$, and we need to know the gluing elements which glue the cells together.

A definition of initial data begins with some number $|\v|$ of particles placed within a two-sphere at the set of points $\{\v\} \in S^2$. Given these points, we choose a CW complex $\Delta$ such that the points $\{\v\}$ define the zero-skeleton $\Gamma^*$. We then specify a rotation parameter\footnote{The rotation parameters must be such that the total mass of the particles is $4\pi$ in order that the spatial hypersurfaces is spherical.} $\bP_\r$ and a bump function $f_\r$ within each region $\r$ of each cell. These three inputs $(\Delta, \bP_\r, f_\r)$ completely specify the initial data.

Once the CW complex, rotation parameters and bump functions $(\Delta, \bP_\r, f_\r)$ have been chosen, we can solve for the rotation functions $a_\r$ as in (\ref{gena}), and also the gluing elements $h_{cc'}$ as in (\ref{h(a)}). Each coordinate function $\z_c$ is defined within its own local Minkowski space via (\ref{genz}). With the rotation and coordinate functions, the fields $(\A, \e)$ are given by (\ref{solutions}) and the particle degrees of freedom can be found by using (\ref{q2}) and (\ref{p1}). Notice that while the rotation and coordinate functions describe flat triangles with rotations supported on the edges only, the field variables $(\A, \e)$ give a different picture where the connection is non-zero within the triangles, and the frame-field has knowledge of this non-trivial $\SU(2)$ connection.

\subsection{Equations of motion}

Given the initial set up described above, let us look at the dynamics generated by the Hamiltonian (\ref{parH}). Since this has support only at the particle locations, we expect the dynamics to be manifest in the motion of the vertices. The fields $(\A, \e)$ describe a triangulation, and motion of the vertices leads to a dynamical triangulation.

Looking at the dynamics generated by the reduced Hamiltonian, we find that the momenta are constants of motion:
\be
\dot{\p^c_\v} = \left\{ \p^c_\v, H \right\} = 0.
\ee
For the positions we calculate:
\be
\left\{(q_\v^c)^i, H \right\} &=& \left\{ (q_\v^c)^i, \alpha_\v \psi_\v \right\} \nonumber \\
&=& \frac{-\alpha_\v}{\sin{\frac{m}{2}}} \left\{ (q_\v^c)^i , \Tr (\p_\v^c) \right\} \nonumber \\
&=& \frac{\alpha_\v}{\sin{\frac{m}{2}}} \Tr \left( \t^i \p_\v^c \right) \nonumber \\
\dot{\q_\v^c} &=& \alpha_\v \u_\v^c.
\ee
where $\u_\v^c$ is the rotation axis associated to $\p_\v^c$. Notice that $\dot{\q}_\v^c = \N_\v(b)$ (where $b$ is the intersection of the $\theta = 1$ line and the boundary $\B_\v$), so that these equations fit with the idea that $\N$ tells us where points on the spatial hypersurface flow under time evolution.

From the viewpoint of a single cell, the above dynamics appears as motion of the vertices and a corresponding change in shape of the triangle. The gluing rules ensure that this dynamics is consistent for all cells which share a vertex. The velocity vector $\dot{\q_\v^c}$ has three components, and it is important that all three vertices of a triangle move at the same rate in the $\ti_c$ direction so that the triangle remains spacelike. This restriction allows us to fix the Lagrange multiplier $\alpha_\v$ by normalizing the $\ti_c$ component of the velocity (i.e. setting $\dot{\q_\v^c} \cdot \ti_c:= -2 \Tr\ (\dot{\q_\v^c} \ti_c) = 1$) with the choice:
\be
\label{alpha}
\alpha_\v = (\u_\v^c \cdot \ti_c)^{-1} .
\ee
Since a neighbouring cell $c'$ which also contains the vertex $\v$ sees a velocity $\dot{\q}_\v^{c'} = h_{cc'}^{-1} \dot{\q}_\v^c h_{cc'}$ and has a basis vector $\ti_{c'} = h_{cc'}^{-1} \ti_{c'} h_{cc'}$, we see that this choice for $\alpha_\v$ normalizes the $\ti_c$ component of the velocity in all cells which contain $\v$.

Under this dynamics, the triangulation is smoothly deformed according to the equations of motion for the fields living on top of the CW complex.
We take the rotation parameters $\bP_\r$ and bump functions $f_\r$ to remain constant under this dynamics. The fields $(\A, \e)$ at any time $t$ are determined by the solutions to the constraints (\ref{solutions}) using the definitions for the rotation (\ref{gena}) and coordinate functions (\ref{genz}); the time-dependence enters in through the coordinate dependence of these equations.

\subsection{Discrete transitions}

We have a nice geometrical picture of particle dynamics within $\mathbb{R} \times \Sigma$. Each cell of the CW space $\Delta$ is given the geometry of a triangle with dynamical vertices according to the definitions for positions and momenta in terms of $(\A, \e)$. The question then is, what happens when a vertex meets an edge, or when two vertices collide? These situations cause discrete changes in the triangulation that affect only some of the cells. We need the particle masses to be maintained in these transition to avoid violating the mass-shell constraints, and we also require the particle data $(\q_\v^c, \p_\v^c)$ in the unaffected cells to be invariant under the transition so that they do not see any change. These conditions single out a unique set of rules for the transitions.

\begin{figure}[ht!]
\centering
\subfigure[Before]{
\includegraphics[scale=0.5]{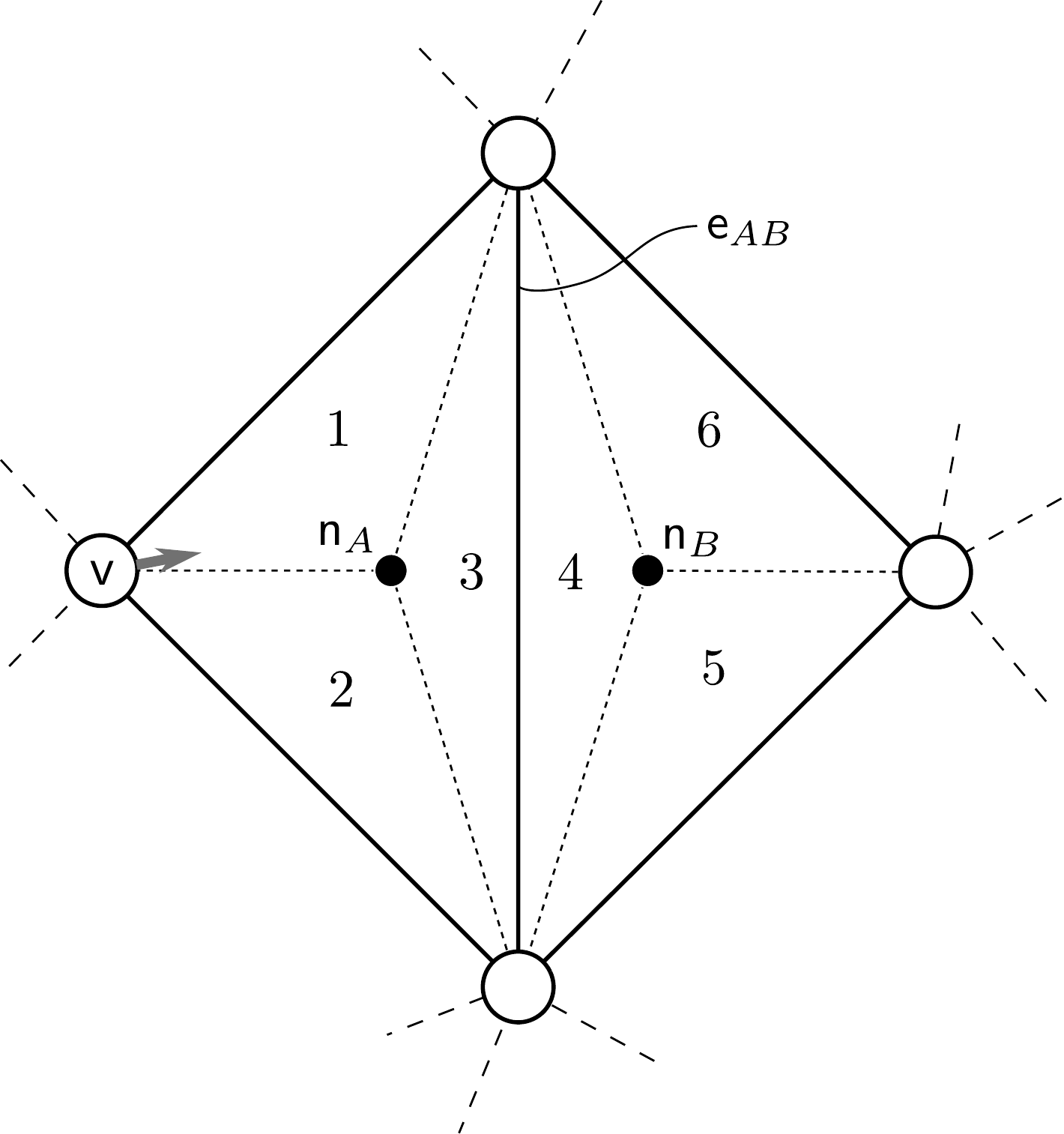}
}
\hspace{1in}
\subfigure[After]{
\includegraphics[scale=0.5]{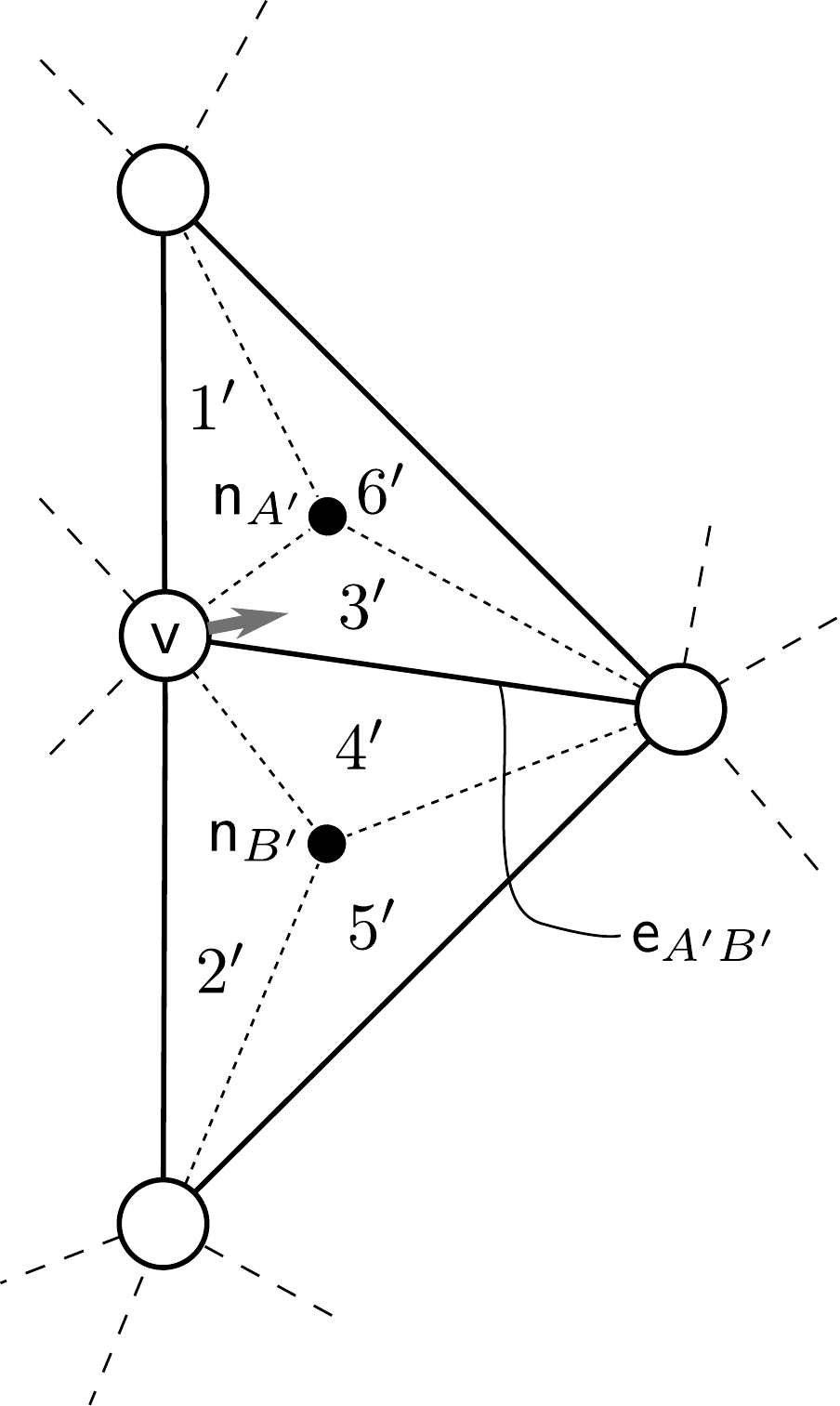}
}
\caption[Bistellar flip]{\label{bistellar} Before and after a discrete transition in the triangulation known as a two-to-two Pachner move, or a bistellar flip. The centroids and the `flipped' edge are labeled, and the numbers label the six regions of the two triangles.}
\end{figure}

Let us first consider the case of a vertex meeting an edge. The general situation which leads to this is depicted in fig. \ref{bistellar}, where the particle $\v$ has a momentum $\p_\v^A$ directed toward the edge $\ed$. When the particle reaches the edge, we define a discrete change in the CW complex and the associated fields as follows. The cell $A$ is removed from the triangulation, while the cell $B$ is split into two along a new edge $\ed_{A'B'}$, i.e. we have a bistellar flip, also called a two-to-two Pachner move. After the transition, we must define new rotation functions within the cells $A'$ and $B'$ by an appropriate choice of the rotation parameters $\bP_\r$. A consistent transition requires:
\be
e^{-\bP_{1'}} = e^{-\bP_1} e^{\bP_3} e^{-\bP_4}, \qquad e^{-\bP_{2'}} = e^{-\bP_2} e^{\bP_3} e^{-\bP_4} \\
\bP_{3'} = \bP_{4'} = 0, \qquad \bP_{5'} = \bP_{5}, \qquad \bP_{6'} = \bP_6,
\ee
while none of the cells outside of these are affected by the transition. Within the new cells $A'$ and $B'$ we can then define the new coordinate and rotation functions, which in turn provide the frame-field and connection within these cells. The total number of triangles is preserved in this transition, as is the total number of particle parameters $(\q_\v^c, \p_\v^c)$, and one can check that the relations (\ref{qrel}, \ref{prel}) continue to hold. After such a discrete transition, the particles again evolve according to the continuous dynamics until another discrete transition occurs.

Let us now check that our definition provides a consistent dynamics. This transition results in a redefinition for six pairs $(\q_\v^c, \p_\v^c) \to (\q_\v^{c'}, \p_\v^{c'})$, since these positions and momenta are now defined within new reference frames. We know that dynamics is a canonical transformation, and one can check that the Poisson algebra is preserved.
As an example, let us check the Poisson bracket for one of the new pairs. The transition rules are such that:
\be
\q_\v^{A'} = h_{AA'}^{-1} \left(\q_\v^{A} - \b_{AA'} \right) h_{AA'}, \qquad \p_\v^{A'} =  h_{AA'}^{-1} \p_\v^A  h_{AA'},
\ee
where $ h_{AA'}$ is the holonomy from $\n_A$ to $\n_{A'}$ at the instant of the transition, and $\b_{AA'}$ is is the translation between these points.
Evaluating the Poisson bracket between the new variables we find:
\be
\left\{ (\q_\v^{A'})^i , \p_\v^{A'} \right\} &=& R(h_{AA'})^i_j \left\{ (\q_\v^{A})^j + \b_{AA'}^j , h_{AA'}^{-1} \p_\v^A  h_{AA'} \right\} \nonumber \\
&=& -R(h_{AA'})^i_j  h_{AA'}^{-1} \t^j \p_\v^A h_{AA'} \nonumber \\
&=& -\t^i \p_\v^{A'},
\ee
where we used that $R(h_{AA'})^i_j \t^j =  h_{AA'} \t^i h_{AA'}^{-1}$ (acting inversely on the basis). Although $\b_{AA'}$ is a covariant integral of the frame field along a path $\gamma$ from $\n_A$ to $\n_B$ (see the footnote 8 at the bottom of page 11), this term does not affect the above Poisson bracket since the path $\gamma$ does not intersect the path $\B_\v$ used to define the momentum $\p_\v^c$. The Poisson bracket between new momenta vanishes trivially, and the final bracket between positions holds by the Jacobi identity. One can check that the Poisson algebra holds for all of the variables after this transition, and that we indeed have a canonical transformation implying this dynamics is well-defined.

\begin{figure}[ht!]
\centering
\subfigure[Before]{
\includegraphics[scale=0.4]{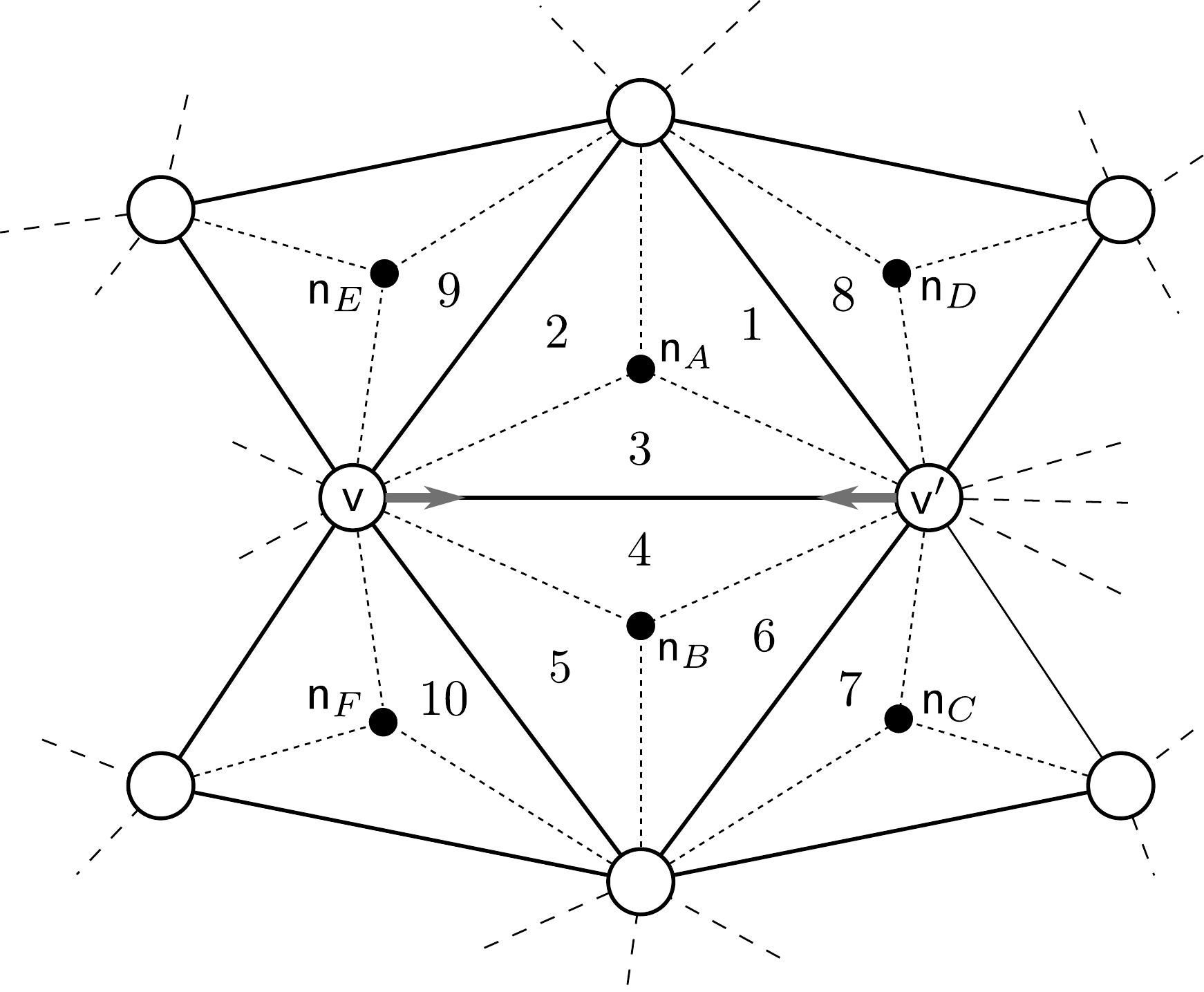}
}
\hspace{0.5in}
\subfigure[After]{
\includegraphics[scale=0.4]{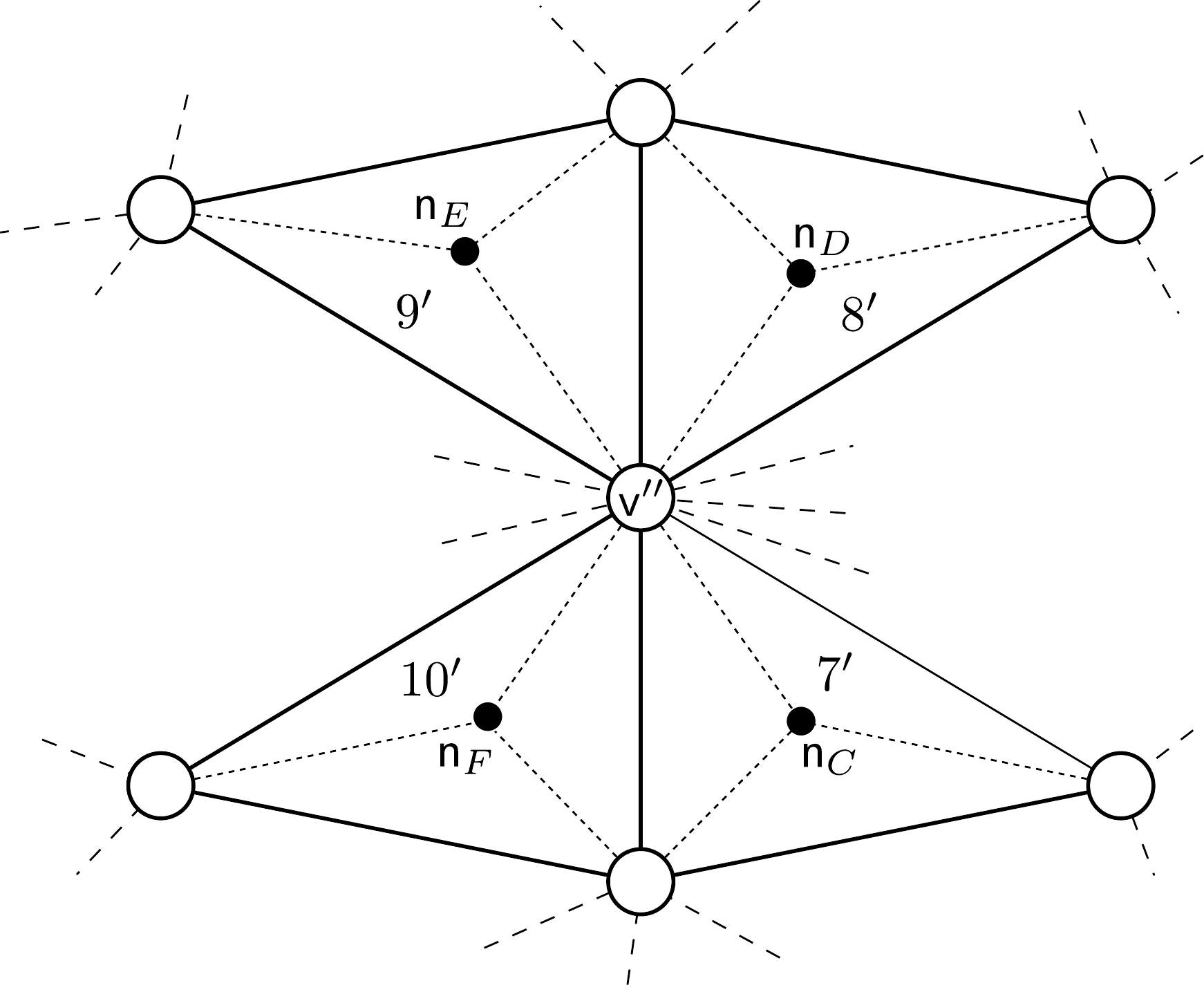}
}
\caption{\label{collide} Before and after a discrete change in the CW complex that occurs when two vertices meet. Upper case roman letters label the nodes (cells), while integers label the relevant regions of the cells. In this transition, the two cells $A$ and $B$ vanish from the CW complex and the vertices $\v$ and $\v'$ combine into one.}
\end{figure}

Let us now consider the case of two vertices colliding as depicted in fig. \ref{collide} where particles $\v$ and $\v'$ are moving toward each other along the edge $\ed_{AB}$. When the particles collide, they stick together and form a new particle $\v''$ whose momentum $\p_{\v''}^c$ is equivalent to the holonomy encircling both particles $\v$ and $\v'$, and the total deficit angle is given by the sum of the masses $m_{\v''} = m_\v + m_{\v'}$. After the collision, both cells $A$ and $B$ are removed from the CW complex and there is a reduction in the physical dimension since two particles have combined to become a single particle. With the following transition rules we have conservation of mass and momenta (as seen from any cell):
\be
e^{\bP_{7'}} = e^{\bP_7} e^{-\bP_6}, \qquad e^{\bP_{8'}} = e^{\bP_8} e^{-\bP_1}, \qquad e^{\bP_{9'}} = e^{\bP_9} e^{-\bP_2}, \qquad e^{\bP_{10'}} = e^{\bP_{10}} e^{-\bP_5}.
\ee
Under this transformation the variables associated to cells $A$ and $B$ are removed, while none of the positions or momenta in cells $C, D, E$ or $F$ are changed, so that this phase space reduction preserves the Poisson algebra of the remaining variables. After such a discrete transition, the particles again evolve according to the continuous dynamics until another discrete transition occurs.

We can now generalize the dynamics discussed in the previous subsection to include discrete transitions. During the continuous evolution of the fields $(\A,\e)$ as the vertices move, the underlying CW complex remains unchanged. When the equations of motion lead to triangle collapsing, we have an instantaneous, discrete transitions in the CW complex $\Delta \rightarrow \Delta'$. Notice that the cells in $\Delta'$ no longer have the same adjacency relations as they did in $\Delta$, so that we in fact have a different triangulation. When and where the discrete transitions occur will of course depend upon our initial choices of $(\Delta, \bP_\r, f_\r)$, and how the dynamics plays out. Let us write this as:
\be
\label{condy}
\begin{array}{cccl}
U(t): & \mathcal{P}^G_{\Gamma^*}  &\longrightarrow    & \mathcal{P}^G_{\Gamma^*} \\
&\Delta            &\longrightarrow & \Delta'  \\
&(\q^c_\v(0), \p^c_\v(0)) &\longmapsto     & (\q^{c'}_\v(t), \p^{c'}_\v(t))  \\
&(\A(x,0), \e(x,0))       &\longmapsto     & (\A(x',t),\e(x',t))
\end{array}
\ee
This evolution is a combination of smooth evolution and discrete transitions according to the rules given above. The process is described entirely within the reduced phase space $\mathcal{P}_{\Gamma^*}$, and we can view this in terms of an evolution in the particle parameters $(\q_\v^c, \p_\v^c)$ which are define in terms of fields $(\A, \e)$.

\subsection{Scattering}
Scattering in three-dimensional gravity can appear to be different for different choices of metric \cite{CCV, BCV, tHooft2}, or in our case, different choices of fields $(\A, \e)$. For instance, the freedom in the initial choice of CW complex will affect when and where the discrete transitions occur. In the DJH model, when and where scattering occurs depends upon where one chooses to cut out the wedges from Minkowski spacetime. In 3d gravity with point particles, there is no well-defined centre of mass frame as used for conventional scattering, and we are stuck with having to choose some other frame to describe the particle trajectories. We note that some work has gone into defining a global metric within which scattering does not depend upon such choices \cite{CCV,BCV}, but still some choice of metric must be made.

In this paper, we have developed a specific choice of geometry that provides a clear picture in terms of evolving triangulations of $S^2$. This setting is not appropriate for discussing the asymptotic trajectories of particles, so we cannot use this to formulate scattering in the conventional sense. However, a gauge-invariant description is possible in terms of so called \textit{particle exchanges}: the action of the braid group on holonomies around the particles \cite{CCV, Carlip}.

Let us consider such an exchange between two particles $\v_1, \v_2$. We fix a base-point $b$ and choose two particle holonomies, i.e. holonomies defined on loops $\gamma_1, \gamma_2$ which go around the corresponding particle and only that particle. The action of the braid group is to wind these holonomies around each other. Following \cite{CCV}, we define a particle exchange operator $\sigma_{12}$ which acts on the tensor space $\mathcal{V}_1 \otimes \mathcal{V}_2$ of $\SU(2)$ holonomies of the two particles. The action of a particle exchange is given by:
\be
\sigma_{12}: \left\{
\begin{array}{rcl}
h_1 & \rightarrow & h_1 \\
h_2 & \rightarrow & h_1 h_2 h_1^{-1}
\end{array}
\right.
\ee
The full monodromy of particle $\v_2$ around particle $\v_1$ is independent of the choice of CW complex. This is given by the action of $\sigma_{12} \sigma_{21}$, i.e. braiding twice.

This picture generalizes to any braiding of an arbitrary number $|\v|$ of particle holonomies $h_{\gamma_\v}$. For a fixed base-point, we can define exchange operators $\sigma_{i, i+1}$ for $i=1, \dots, |\v|-1$ which act on the tensor space $\mathcal{V}_1 \otimes \cdots \otimes \mathcal{V}_V$ of particle holonomies to generate the braid group $B_V$. The braid group provides a useful tool for understanding the Jones polynomial of knot theory \cite{Witten} and plays an important role in the quantization of three-dimensional gravity with point sources (see the papers by Freidel and Louapre listed in \cite{FL}). In this classical setting, the discrete transitions described above can be written in terms of the action of generators $\sigma_{i, i+1}$ on particle holonomies. One can check this by choosing a base point and writing the holonomies around particles in terms of the rotation functions $a_\r$ associated to regions. As particles move, one finds that these holonomies can change under the discrete transitions according to the braiding action given above. Note that whether or not a single braid appears during a discrete transition depends upon the choice of base-point. Only once a particle has gone completely around the other do we get a result that does not depend upon the choice of base-point.

This concludes our analysis within the framework of general relativity. We have given the particle data in terms of a position and momentum $(\q_\v^c, \p_\v^c)$ for each particle, in the frame of each cell that contains the particle. This geometrical data is defined within a topological space known as a CW complex, and the geometry associated to each cell is that of a triangle. Masses are found from the norm $m_\v = |\p_\v|$. Dynamics leads to motion of the vertices which causes the triangles to change shape. It may happen that a vertex reaches an edge or another vertex causing a discrete change in the underlying topology and a redefinition of the geometrical data. A description of trajectories depends upon how one chooses the CW complex and fields $(\A, \e)$, but a gauge invariant description of scattering is provided by the braid group. 

Remarkably, the system we have described here admits a description in terms of the loop gravity phase space. In the next section we define this phase space and develop an equivalent description of the system in terms of the loop gravity variables.

\section{Point particles in the loop gravity phase space}

Now that we have established the theory of 3d point particles in the more conventional setting of general relativity, let us develop this theory in the framework of loop gravity. We begin by introducing the loop gravity phase space in the following subsection.

\subsection{Loop gravity phase space}
The continuous phase space $\mathcal{P}$ that we used above to describe general relativity is parameterized by fields $(\A, \e)$ defined upon a CW complex. Rather than fields, the discrete loop gravity phase space $P_\Gamma$ is parameterized by a finite number of parameters $(h_\l, \X_\l)$ for $\l = 1, \dots, |\l|$. Instead of a CW complex, the parameters $(h_\l, \X_\l)$ are defined upon an oriented graph $\Gamma$ as shown in fig. \ref{graph}. A different choice of graph will lead to a different phase space (by definition), although the same physical data may be described using different graphs.

Let us now be more precise about the definition of the loop gravity phase space $P_\Gamma$. We shall focus on the two-dimensional case required for the model at hand in order to keep the presentation simple. In two dimensions, we begin with a choice of abstract, oriented graph $\Gamma$ as shown in fig. \ref{2dgraph}. `Oriented' implies that each link has a direction, and `abstract' means that the graph is not embedded within any space. These graphs are topological spaces and their shape has no meaning; the only information here is the number of links, their orientations, and how they are connected at nodes. Each node of the graph must connect at least three links, and we do not consider any graphs which have knotted links.
\bcf
\includegraphics[width=0.4\linewidth]{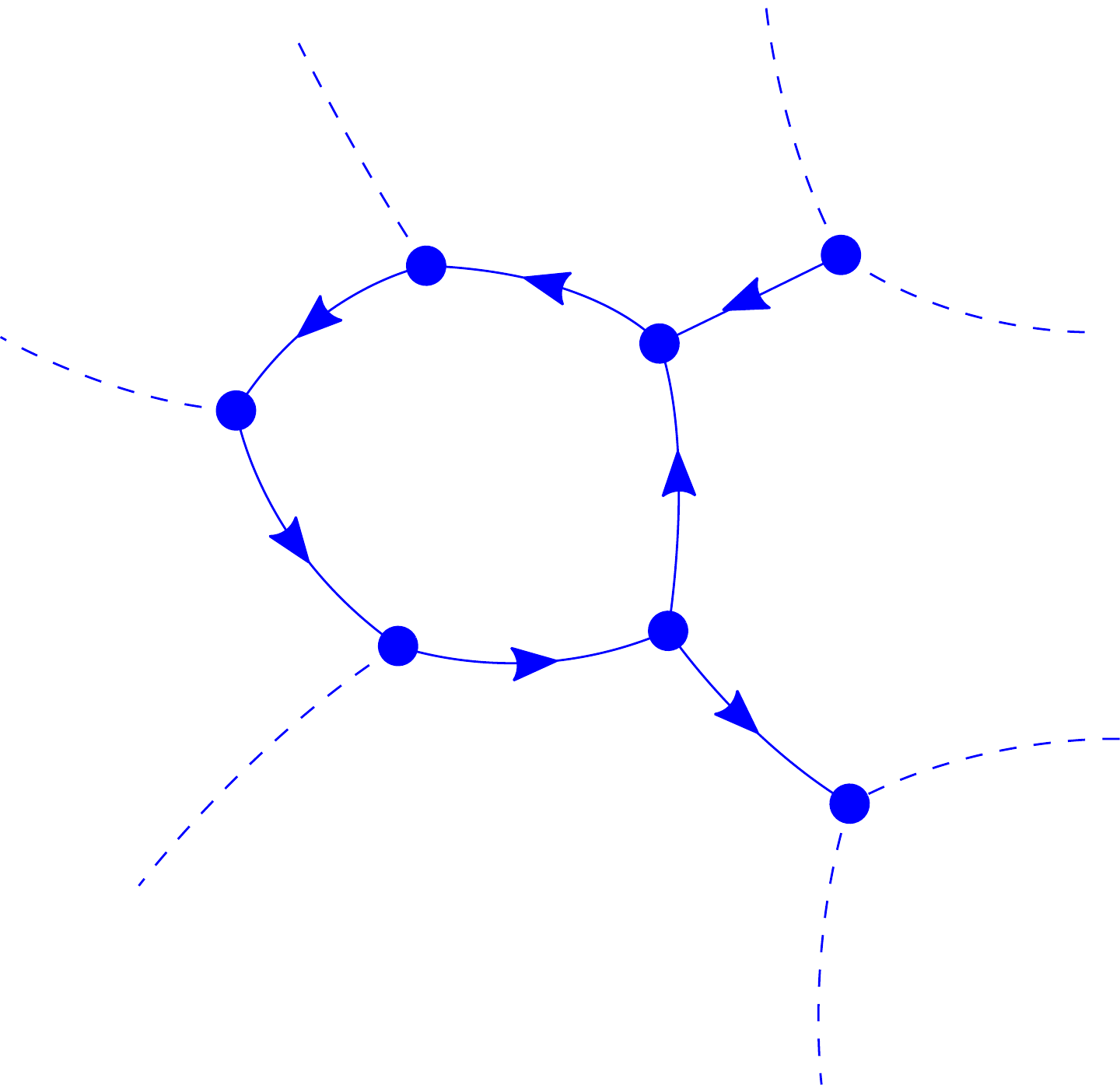}
\caption{\label{2dgraph}Neighbourhood of an oriented graph. Links and vertices are shown, with an arrow on each link to indicate orientation. Some links are drawn with dashes to show where the neighbourhood joins with the graph as a whole.}
\ecf

Given an abstract, oriented graph (hereafter just `graph'), we assign a pair of variables to each link. The first of these is the holonomy $h_\l \in \SU(2)$ which provides a notion of parallel transport along the link. The second variable is the \textit{flux} $\X_\l \in \su(2)$, which in two dimensions is a vector with units of length. Each pair $(h_\l, \X_\l)$ parameterizes the cotangent bundle $T^*\SU(2)$, and the phase space over the entire graph is obtained by taking the direct product over all of the cotangent bundles:
\be
P_\Gamma \equiv \underset{\l}{\times} T^* \SU(2)_\l .
\ee
The variables $(h_\l,X_\l)$ satisfy the Poisson algebra:
\be\label{poisson algebra2}
\big\lbrace \X^i_\l,\X^j_{\l'}\big\rbrace=\delta_{\l \l'}\epsilon^{ij}_{~~k}\X^k_\l,\qquad\big\lbrace \X^i_\l,h_{\l'}\big\rbrace=-\delta_{\l \l'}\t^i h_\l,\qquad\big\lbrace h_\l,h_{\l'}\big\rbrace=0 .
\ee

As for the continuous theory, we can define a Gauss constraint in terms of these discrete variables which generates $\SU(2)$ gauge transformations. In fact, one can derive the discrete Gauss constraint from the continuous one \cite{FGZ}. The discrete Gauss constraint is defined at each node $\n$ by:
\be
G_\n = \sum_\l \X_\l .
\ee
Under this constraint, the fluxes associated to a node must sum up to zero. Since fluxes represent length vectors, this implies that the fluxes associated to a single node represent the edge vectors of a polygon. For this reason the Gauss law is also called the closure constraint in this discrete context.

The discrete Gauss constraint generates gauge transformations. Given an element $g_\n \in \SU(2)$ at each node $\n$, the finite gauge transformations are given by:
\be\label{SU(2)}
h_\l \to g_{s(\l)}h_\l g_{t(\l)}^{-1},\qquad\qquad \X_\l \to g_{s(\l)}\X_\l g_{s(\l)}^{-1},
\ee
where $s(\l)$ (resp. $t(\l)$) denotes the starting (resp. terminal) node of $\l$.

By taking the discrete Gauss constraint into account, we can define the gauge-invariant phase space:
\be
P^G_\Gamma=\underset{\l}{\times}T^*\SU(2)_\l\sslash\SU(2)^{|\n|} ,
\ee
by symplectic reduction where $|\n|$ is the number of nodes in the graph. The double quotient means to impose the Gauss constraint at each node $\n$ and divide out the action of the $\SU(2)$ gauge transformations (\ref{SU(2)}) that it generates, i.e. to identify values of the parameters which are related by $\SU(2)$-gauge transformations.

\subsection{Relating the continuous and discrete phase spaces}

The loop gravity phase space can be related to the gravitational phase space by embedding a graph within a CW complex. Given a CW complex $\Delta$, there is a particular graph which is its dual, and vice versa. For such a dual pair, there is a one-to-one correspondence between nodes of the graph and cells of the $\Delta$, as well as links of the graph with edges of $\Delta$. To embed a dual graph within a CW complex, we place a node $\n$ at the centroid of each triangle, and choose the links $\l$ between these nodes so that there is one link in the graph intersecting each edge of the CW complex. See fig. \ref{triangulation} for an illustration.
\bcf
\includegraphics[width=0.4\linewidth]{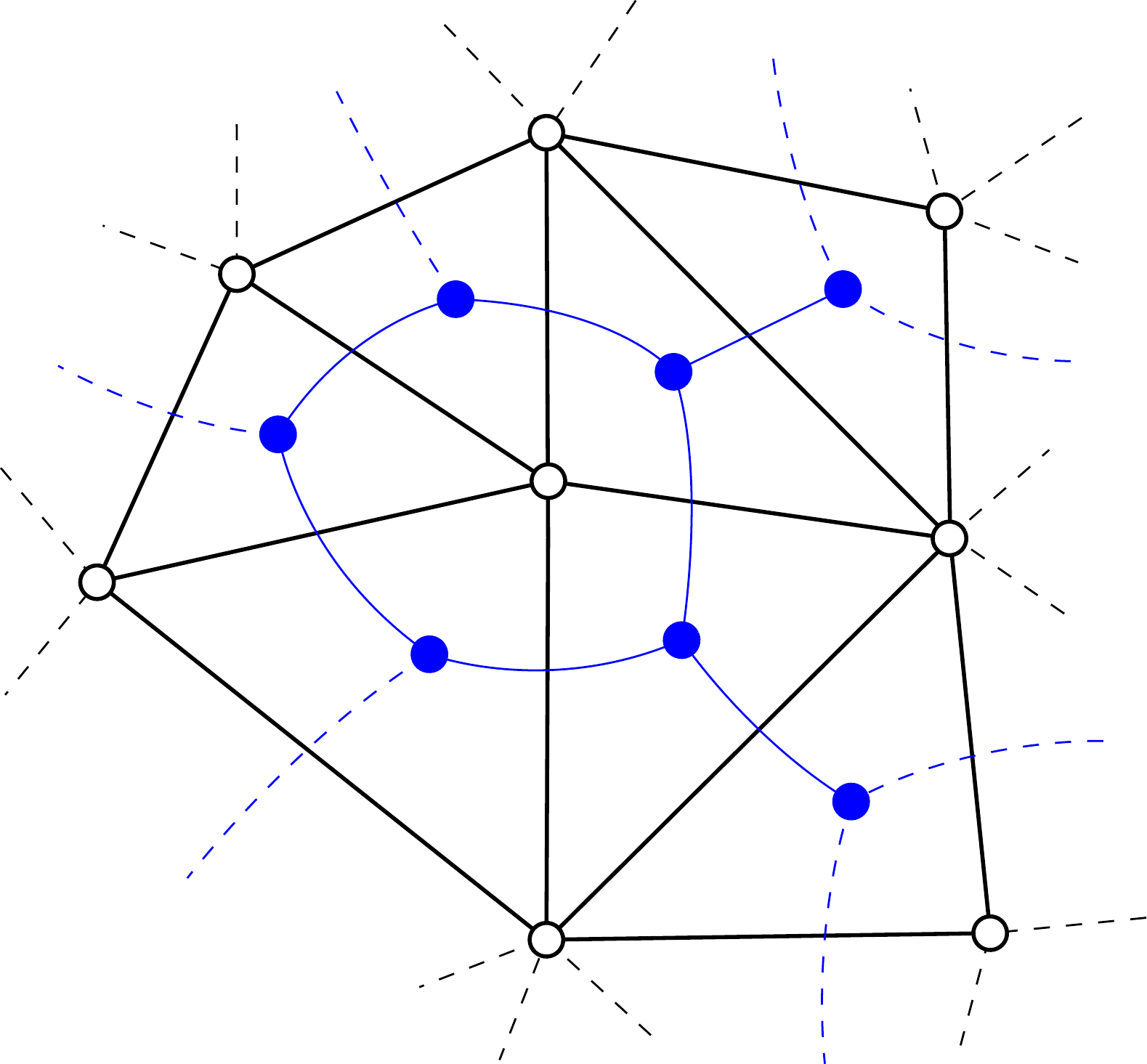}
\caption{\label{triangulation}Neighbourhood of a CW complex and its dual graph. Edges (black lines) and vertices (open black circles) of the CW complex are shown, together with a dual embedded graph $\Gamma$ consisting of links (blue curves) and nodes (filled blue circles).}
\ecf

With a graph embedded within a CW complex, we now give the map from an arbitrary connection to the holonmies. For a link $\l_{cc'}$ from the node $\n_c$ to $\n_{c'}$ we have:
\be
\label{hdef}
h_{cc'} = \Pexp \int_{\l_{cc'}} \A ,
\ee
while the holonomy for the reversed link $\l_{cc'}^{-1} = \l_{c'c}$ is given by the inverse:
\be
\label{hrel}
h_{c'c} = h_{cc'}^{-1}.
\ee
If we use the particular choice of fields $(\A,\e)$ described in the gauge fixing procedure above, we can write the holonomies in terms of rotation parameters:
\be
h_{cc'} = a_\r^{-1}(x) a_{\r'}(x)\left.\right|_{\ed_{cc'}} = e^{\bP_\r} e^{-\bP_{\r'}},
\ee
where the regions $\r \in c$ and $\r' \in c'$ are on either side of the edge, and the rotation functions are evaluated at the edge.

We define the fluxes as an integral of the frame-field $\e$, although the connection also enters here in order to make the integral covariant. For general fields $(\A, \e)$, the flux associated to the link $\l_{cc'}$, as seen from the node $\n_{c}$, is given by:
\be
\label{fluxdef}
\X_{cc'} = \int_{\ed_{cc'}} h_\pi \e h_\pi^{-1} ,
\ee
where $\pi$ is a set of paths from $\n_c$ to the points of integration. Let us call the endpoints of the edge $\v$ and $\v'$. Then, for our gauge choice of the fields $(\A, \e)$, we can write the flux in terms of coordinate functions only:
\be
\X_{cc'} = \int_{\ed_{cc'}} h_\pi \e h_\pi^{-1} = \int_{\v}^{\v'} \rd \z^c =  \z^c (\v ') - \z^c (\v) ,
\ee
where $h_\pi(x) = a_\r^{-1}(x)$. This result is independent of the choice of paths since the rotation functions depend on $\theta_\r$ only and are constant along the edge. Notice this is just the relative distance between endpoints of the edge, as seen from the point $\n_c$.
It is easy to see that the fluxes associated to a node satisfy the closure constraint (Gauss law):
\be
\label{disGauss}
\sum_{\l \ni \n} \X_\l = 0,
\ee
since this is an integral around a closed loop.
These fluxes also satisfy the gluing relation:
\be
\label{Xrel}
\X_{c'c} = \int_{\ed_{c'c}} h_\pi \e h_\pi^{-1} = \z^{c'} (\v) - \z^{c'} (\v') =  - h_{cc'}^{-1} \X_{cc'} h_{cc'}.
\ee
This is an expression of the relative distance between $\v'$ and $\v$ as seen from the cell $c'$.

In general, the definitions (\ref{hdef}, \ref{fluxdef}) provide a non-invertible map from the phase space $\mathcal{P}$ to the discrete space $P_\Gamma$. However, if we impose the flatness and Gauss constraints then these definitions provide an invertible (one-to-one) map from $\mathcal{P}_{\Gamma^*}^G$ to the discrete, gauge-invariant phase space $P_\Gamma^G$ \cite{FGZ}. This implies that the two phase spaces are isomorphic and describe the same physics, so we can expect to fully describe 3d point particles in the loop gravity framework.

We commented previously that there are different choices of $\Delta$ for a set of points $\Gamma^*$, and that these choices are related by two-to-two Pachner moves. Each choice of CW complex leads to a different dual graph, and these different graphs are also related by two-to-two Pachner moves. What these choices amount to are different fields $(\A, \e)$ on the continuous side, or different sets of $(h_\l, \X_\l)$ on the discrete side. On the continuous side, this is a gauge choice since any choice yields the same physical results. On the discrete side however, this is different than a gauge choice since a change in graph that gives a new phase space by definition. This implies that different graphs can lead to phase spaces which contain the same physical data.

\section{Dynamics on the graph}
In order to develop the loop gravity dynamics, we shall now write the reduced Hamiltonian (\ref{H_R}) in terms of loop variables. This is just a sum over the mass-shell constraints, which are written in terms of traces of the holonomies circling each particle. Now, the holonomies $h_\l \in P_\Gamma$ associated to the graph we have embedded go from the centroid of one triangle to the next, and none of these lie on the particle boundaries. However, these constraints are independent of choice of base-point since this data is washed out in the trace. Also, since the connection is flat the choice of path does not matter and we can circle each particle going from centroid to centroid, using the $h_\l$ data we have at hand.

We have already written the Hamiltonian in terms of particle momenta in the previous section. This is easily translated into the loop variables since the particle momenta (\ref{p1}) are given in terms of the gluing elements $h_{cc'}$, which are equivalent to the holonomies $h_{\l}$ on the graph. We have:
\be
H_R = \sum_\v \alpha_\v \psi_\v, \qquad \psi_\v := 2 \cos^{-1} \frac{W_\v}{2} - m_\v ,
\ee
where the Wilson loop around $\v$ is:
\be
W_\v = \Tr (\prod_{\l \in \o_\v} {h_\l}),
\ee
where $\o_\v$ is a loop composed of the links encircling $\v$, the product is ordered counterclockwise around $\v$, and the choice of base-point has no consequence.
Through these constraints, we have the mass of each particle $\v$ in terms of the holonomies $h_\l \in P_\Gamma^G$. The question now is: What evolution does this Hamiltonian generate on the graph data?

Let us define a set of initial data for the dynamics in the loop gravity picture. We first require a graph that is dual to a CW complex $\Delta$, so that each node must connect three links. Upon each link of the graph, we choose a consistent set of holonomies and fluxes that satisfy the relations between cells (\ref{hrel}, \ref{Xrel}) and the closure constraint (\ref{disGauss}). In addition, the total mass of all particles must equal $4\pi$ so that the graph is dual to a triangulation of $S^2$. This is our initial data.

The holonomies are constants of motion since:
\be
\dot{h}_\l = \left\{ h_\l, H_R \right\} = 0.
\ee
The interesting dynamics is seen in the fluxes. These represent the edge vectors of the triangulation, and changes in the flux correspond to changes in the triangulation. The Hamiltonian is written in terms of a product of holonomies around each particle $\v$. Where $m$ cells meet at a particle, let us label the cells such that the loop starts at the node $\n_1$. From the holonomy-flux Poisson algebra, we have for any cells $c$ and $c+1$ in the loop $\o_\v$:
\be
\left\{X^i_{c,c+1}, W_\v \right\} &=& \left\{X^i_{c,c+1}, \Tr \  h_{1, 2} h_{2, 3} \cdots h_{m-1, m} h_{m, 1}  \right\} \nonumber \\
&=&  \Tr \ h_{1 2}\cdots h_{c-1, c} (-\t^i) h_{c,c+1}\cdots h_{m, 1}  \nonumber \\
&=& - \Tr \ \t^i h_{c,c+1}\cdots h_{m, 1} h_{1 2}\cdots h_{c-1, c} \nonumber \\
&=& -\sin(\frac{m_\v}{2}) \left( u^c_\v \right)^i
\ee
If the link $\l_{c, c+1}$ is not part of the loop $\o_\v$ the bracket vanishes. Recall that $\left( u^c_\v \right)^i$ is a unit vector in the direction of particle momentum as seen in the cell $c$.
Using this bracket we can find the equation of motion for the flux associated to the edge between two vertices $\v$ and $\v'$:
\be
\left\{X^i_{c,c+1}, H_R \right\} &=& \left\{X^i_{c,c+1}, \sum_\v \alpha_\v 2 \cos^{-1} \frac{W_\v}{2} \right\} \nonumber \\
&=& \frac{-\alpha_\v}{\sin{\frac{m}{2}}} \left\{ X^i_{c,c+1} , W_\v \right\} + \frac{\alpha_{\v'}}{\sin{\frac{m}{2}}} \left\{ X^i_{c,c+1} , W_{\v'} \right\} \nonumber \\
\dot{\X}_{c,c+1}&=& \alpha_\v \u^c_\v - \alpha_{\v'} \u^c_{\v'},
\ee
where the $\alpha_\v$ are normalization constants given by (\ref{alpha}). A flux is the relative distance between two vertices, and as one might expect, the equation of motion is the difference between the equations of motion for the endpoints.

Since each set of three fluxes $\X_\l$ intersecting at a node defines a triangle, the evolution of the loop gravity variables on a graph describes a time dependent triangulation just as we had in the framework of general relativity. In fact, the loop gravity variables describe the \textit{same} triangulation since the fluxes are given by the differences between positions $\q_\v^c$. The $\SU(2)$ rotations $h_\l$ remain constant, as did the gluing elements $h_{cc'}$. Given some initial triangulation defined by the data $(h_\l, \X_l)$, the triangulation at some later time is given by a smooth diffeomorphism achieved by shifting the vertices.

\subsection{Discrete transitions of the graph}

When we looked at the dynamics in terms of $(\q_\v^c, \p_\v^c)$, we found that a bistellar flip occurs whenever a vertex moved onto an edge. In terms of fluxes, this happens whenever the fluxes associated to a single node become parallel ($[\X_\l, \X_{\l'}] = 0$), with none of the fluxes vanishing. We need to find the corresponding rules which describe this transition in terms of $P_\Gamma^G$ phase spaces. These follow directly from the definitions we gave for the triangulation.

A bistellar flip in the triangulation corresponds to a two-to-two Pachner move on the dual graph.
\begin{figure}[ht!]
\centering
\subfigure[Before]{
\includegraphics[scale=0.5]{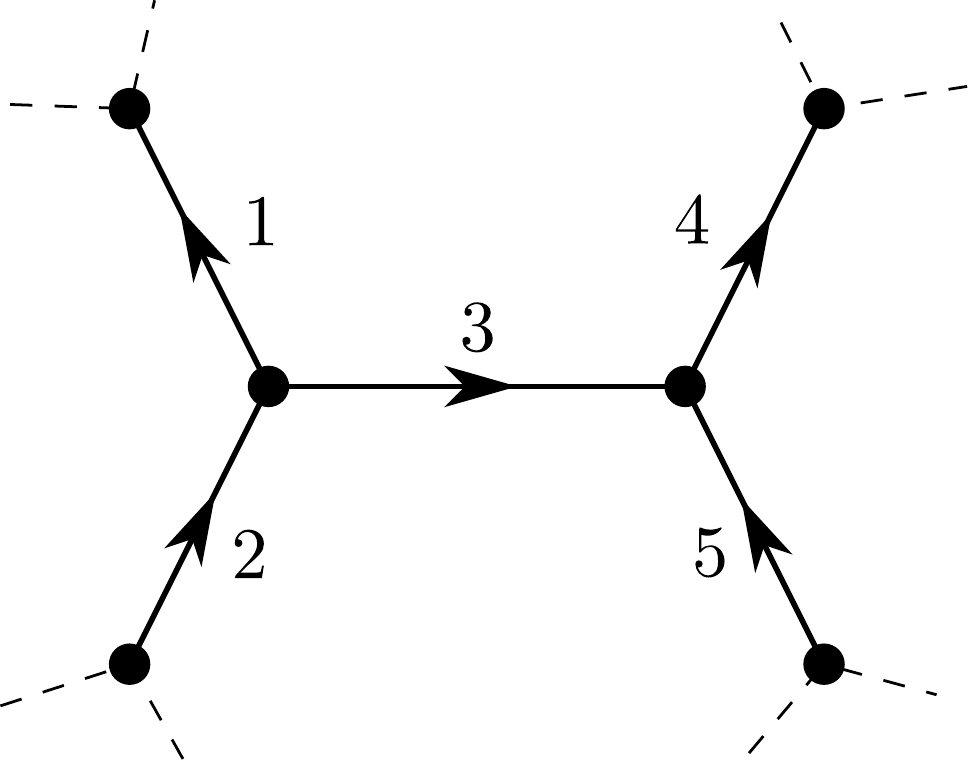}
}
\hspace{1in}
\subfigure[After]{
\includegraphics[scale=0.5]{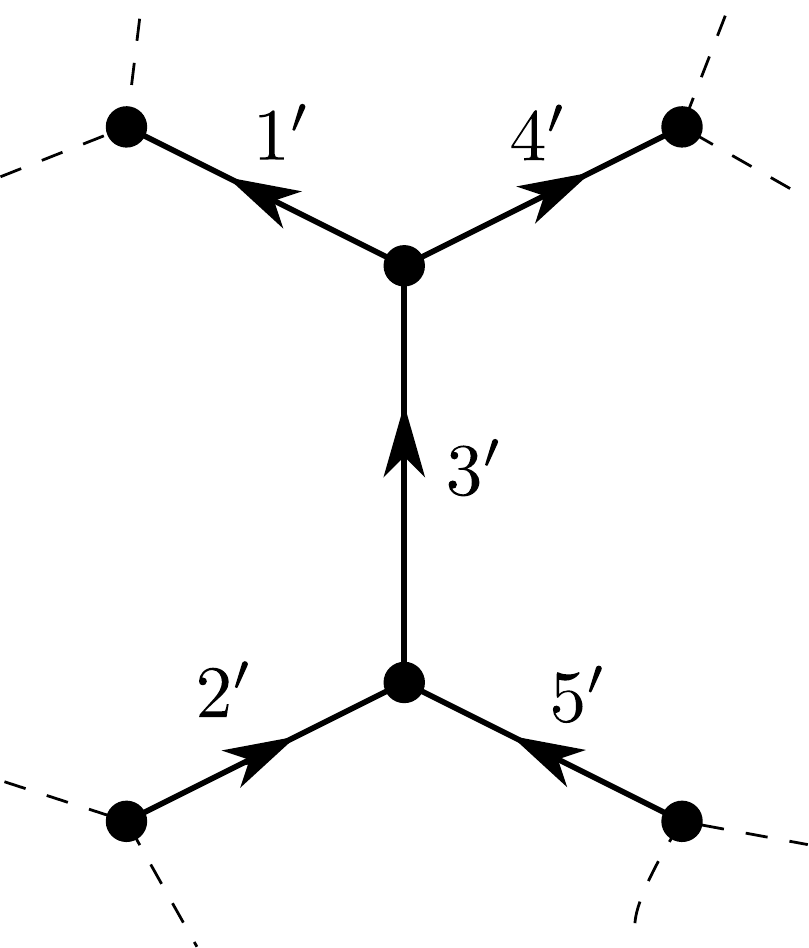}
}
\caption[Two-to-two Pachner move]{\label{Pachner} Before and after a discrete transition in the graph known as a two-to-two Pachner move. The links are each labeled with a number.}
\end{figure}
Consider the holonomies as labeled in fig. \ref{Pachner}, under the transition that occurs when fluxes $(\X_1, \X_2, \X_3)$ become parallel while they each maintain a finite length. This is dual to the transition shown in fig. \ref{bistellar} that we discussed above for the continuous formulation. In order for a consistent duality between the discrete and continuous pictures, we have for the holonomies that:
\be
h_{1'}= h_{3}^{-1} h_1, \qquad h_{2'} = h_2 h_3, \qquad h_{3'} = \mathbb{1}, \qquad h_{4'} = h_4, \qquad h_{5'} = h_5.
\ee
This definition preserves the holonomies around particles $h_{\o_\v, n_c}$ for each node except the two which are attached to $\l_3$. These nodes are replaced by new nodes that are dual to the new triangles. For the fluxes we have:
\be
\X_{1'} = h_{3}^{-1} \X_1 h_3, \qquad \X_{2'}^{-1} = h_{3}^{-1} \X_2^{-1} h_3, \qquad \X_{4'} = \X_4, \qquad \X_{5'} = \X_5.
\ee
In order to determine $\X_{3'}$ we use the closure constraint, taking orientations into account:
\be
\X_{3'} = -\X_{2'}^{-1} - \X_{5'}^{-1} = -h_{3}^{-1} \X_2^{-1} h_3 - \X_5^{-1}.
\ee
One can check that the relationships (\ref{hrel}, \ref{Xrel}) and Gauss law (\ref{disGauss}) remain consistent under this transition.

This transition takes us from the phase space associated to the graph $\Gamma$, to a different phase space associated to a new graph $\Gamma'$. However, the number of degrees of freedom are preserved, and this transition is a canonical transformation. We can check to see that the new variables satisfy the Poisson algebra of $T^*(\SU(2))$. Let us check explicitly for $h_{1'}$ and $\X_{1'}$ as an example. We have trivially that:
\be
\left\{ h_{1'}, h_{1'} \right\} = 0.
\ee
For the fluxes we find:
\be
\left\{ \X_{1'}, \X_{1'} \right\} &=& \left\{  h_{3}^{-1} \X_1 h_3,  h_{3}^{-1} \X_1 h_3 \right\} \nonumber \\
&=& h_3^{-1} [\X_1, \X_1] h_3  \nonumber \\
&=& [\X_{1'}, \X_{1'}] ,
\ee
as desired, and the final bracket also checks out:
\be
\left\{ X_{1'}^i, h_{1'} \right\} &=& R(h_{3})^i_j \left\{ \X_1^j ,  h_{3}^{-1} h_1 \right\} \nonumber \\
&=& - R(h_{3})^i_j h_{3}^{-1} \t^j h_1 \nonumber \\
&=& - \t^i h_{3}^{-1} h_1 \nonumber \\
&=& - \t^i h_{1'}
\ee
Note that the holonomy $h_{3'}$ is not given in terms of holonomies on the `before' graph, but has been assigned to the identity. Because of this, we cannot use the `before' Poisson brackets to define the `after' Poisson brackets for this link. We must assign the $T^*SU(2)$ algebra to the variables on link $\l_{3'}$ in order to be consistent with the mapping from the continuous fields $(\A, \e)$ within the new dual cells. With this definition, we then have that this transition is a canonical transformation between the variables in $P^G_\Gamma$, to the variables parameterizing $P^G_{\Gamma'}$ on a new graph.

\begin{figure}[ht!]
\centering
\subfigure[Before]{
\includegraphics[scale=0.5]{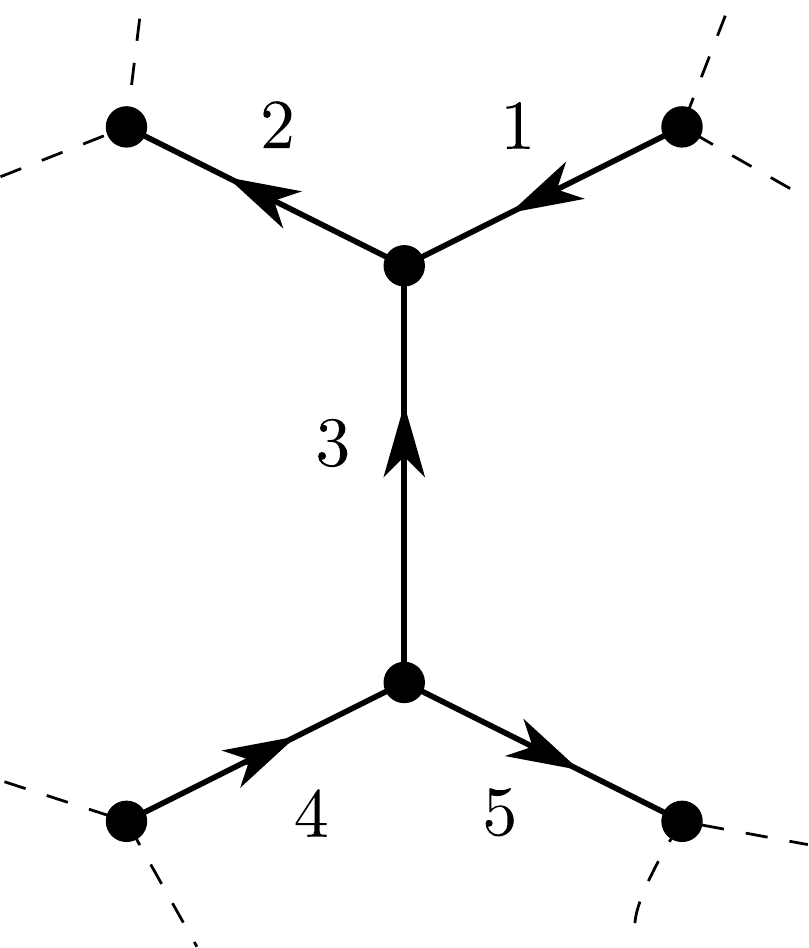}
}
\hspace{1in}
\subfigure[After]{
\includegraphics[scale=0.5]{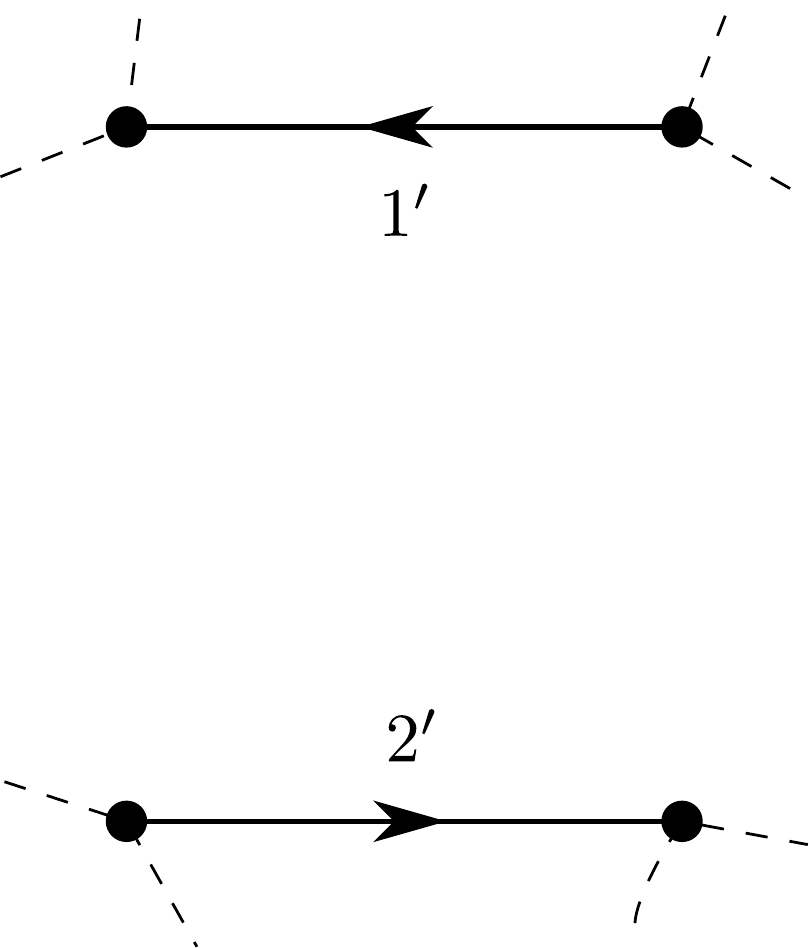}
}
\caption{\label{collidegraph} Before and after a discrete change in the graph that occurs when the flux $\X_3$ vanishes. Each link is labeled by an integer.}
\end{figure}

Let us now look at what happens when one of the fluxes vanishes, implying that two vertices have collided and become one as shown in fig. \ref{collidegraph}. Notice that when a flux vanishes, the remaining two fluxes become equal in magnitude while their relative signs are fixed by the orientation of the link. This case is rather simply described in the loop gravity framework. Here, when the flux $\X_3$ vanishes, the link $3$ is removed from the graph while the composition of link $1$ with link $2$ becomes a single link labeled $1'$, and the composition of links $4$ and $5$ becomes the link $2'$. We have the following rules for determining the new variables:
\be
h_{1'} = h_1 h_2, \qquad h_{2'} = h_{4} h_5 , \qquad \X_{1'} = \X_{1} = \X_2, \qquad \X_{2'} = \X_4 = \X_5.
\ee

In the loop gravity picture, we see dynamics in terms of changes in the flux associated to the relative distances between particles. This is not amenable to a description of particle scattering in terms of trajectories. However, the action of the braid group is easily given in terms of the holonomies $h_\l$ associated to the links on the graph, so the discussion of braiding given above for the continuous phase space applies here as well.

Taking the discrete transitions of the graph into account, the equations of motion for $(\X_\l, h_\l)$ describe the dynamics $U_{\Gamma \Gamma'}(t)$ of a moving triangulation which we write as:
\be
\begin{array}{cccl}
U_{\Gamma \Gamma'}(t): & P_\Gamma^G &\longrightarrow       & P_{\Gamma'}^G                   \\ 
&\Gamma            &\longrightarrow       & \Gamma'                         \\
&(h_\l(0), \X_\l(0))     &\longmapsto           & (h'_{\l'}(t), \X'_{\l'}(t))
\end{array}
\ee

We have found a dynamics in terms of holonomies and fluxes on a graph $\Gamma$ that is consistent with the evolution of continuous fields $(\A, \e)$ on a CW complex $\Delta$. The data $(h_\l, \X_\l)$ describes a triangulation and the equations of motion dictate how the triangulation moves. Consistency with the continuous picture tells us how the graph changes when a triangle collapses. The change in graph causes a change of phase space at the instant of the transition, although the `after' phase space describes the same physics as the `before' phase space. There are discrete changes given by two-to-two Pachner moves which preserve the number of links on the graph, and therefore also preserves the dimension of the phase space. When one of the fluxes vanishes, we lose a link of the graph and have a corresponding reduction in dimension as two of the particles have joined into one.

\section{Conclusion}
Point particles in 2+1 dimensional Riemannian gravity make a nice test theory for loop classical gravity. We first developed this theory within the framework of general relativity to help guide us in the loop gravity formulation, and also to ensure that these different formulations are in agreement. The continuous phase space is given by a connection and frame-field $(\A, \e) \in \mathcal{P}$, each taking values in $\su(2)$. There is a flatness constraint that arises naturally from the Hamiltonian decomposition of the action, which restricts the curvature $\F(\A)(x) = 0$ for all $x \ne \Gamma^*$. Point particles reside at the locations $\v \in \Gamma^*$ where curvature is supported. In addition to the flatness constraint, there is a Gauss constraint, which on the 2d hypersurface is equivalent to a zero-torsion condition. We showed explicitly the gauge fixing procedure which selects a representative geometry $(\A, \e)$ and allows us to solve for the Lagrange multipliers. The geometry defined by the gauge choice is the 2d analog of spinning geometries \cite{FZ}, which also have $\F(\A) = \rd_A \e = 0$ everywhere within the cells.

After the gauge-fixing procedure, we are left with a reduced Hamiltonian given by a sum of mass shell constraints, determined by holonomies around the particles. The reduced Hamiltonian can be written in terms of a position and momentum $(\q_\v^c, \p_\v^c)$ for each vertex $\v$ and each frame $c$ which contains the vertex. This allows us to describe the dynamics in terms of an evolving triangulation where the fields $(\A, \e)$ evolve continuously within a fixed topological space. We are able to define how the geometry and the underlying CW complex changes when a triangle collapses. On the other hand, we used the isomorphism between $P_\Gamma^G$ and the reduced phase space $\mathcal{P}_{\Gamma^*}^G$ to study this system in terms of the holonomies and fluxes on a graph. We are able to use the holonomy-flux equations of motion to define the evolution of the same triangulation by embedding a dual graph $\Gamma$ with the CW complex $\Delta$. This allows us to define discrete transitions on the graph such that the duality is maintained, and the evolving geometry described by the discrete framework coincides with that described by the continuous framework.

The main result here is that we have described a gravitational system entirely within the loop gravity framework, and that this description agrees precisely with that given in terms of fields for general relativity. The loop gravity description is well-suited for quantization by the established methods of LQG, and since the classical theory agrees with general relativity, the quantization would yield a 3d quantum theory of general relativity. It would be very interesting to carry out this quantization and compare results with the spin foam models point 3d point particles in \cite{FL}.

Our work on this toy model has uncovered some features that we might expect to carry over to the full theory. Discrete graph changes are expected by many to be necessary for describing gravitational wave propagation in four dimensions \cite{Smolin}, and here we have an explicit realization of this. We have also developed some unconventional ideas which may help to advance the 4d theory. Here we treated the graph as a topological object rather than embedding it within some geometry. The dynamics of LQG is generally expected to be given entirely in terms of discrete evolution moves, but here we found that the classical evolution is continuous, but with intermittent discrete changes. With these hints from 3d gravity, we can continue working toward a 4d theory of classical loop gravity which agrees with general relativity. There is a major technical leap in going to four spacetime dimensions, since the phase space for gravity is then infinite-dimensional as opposed to the finite-dimensional phase spaces we dealt with here. It is a very difficult problem to understand how to describe these infinite degrees of freedom in terms of finite dimensional phase spaces $P_\Gamma$. However, \textit{if} this can be done, then applying the already well-established methods of LQG would lead toward a full theory of quantum general relativity! This is a big `if', but the point here is that the difficulties facing LQG may not be in the quantization, but rather in describing gravity in terms of the holonomies and fluxes on graphs. This direction of research is designed to single out this problem and attack it.

\section*{Acknowledgments}
I would like to thank Gabor Kunstatter and Laurent Freidel for numerous discussions throughout the course of this work. I also thank Jack Gegenberg, Viqar Husain and Sanjeev Seahra for helping me to clarify these ideas in the final stages of this project.

\appendix
\section{Holonomy around a particle boundary $\B$}
\label{appendix}
In this appendix we review some relevant properties of holonomies and derive an expression for the holonomy around the particle boundary $\B$. Excising a particle from the spacetime leaves a smeared cylindrical boundary around the particle worldline, and we refer to $\B$ as the intersection between this boundary and the spatial hypersurface $\Sigma$.

The geometric meaning associated to a holonomy $h_\gamma \in \SU(2)$ is the parallel transport of a vector along a path in spacetime. Recall that an element $\bm{\xi} \in \su(2)$ is associated with a vector through the identification $\xi^i = -2 \Tr (\bm{\xi} \bm{\tau}^i)$. Under parallel transport along a curve $\gamma$, the vector transforms as:
\be
\bm{\xi} \rightarrow h_\gamma \bm{\xi} h_\gamma^{-1}.
\ee

We may parameterize a path (that does not intersect itself) as $\gamma(s)$, where $s$ takes values over the interval $[0,1]$:
\be
\gamma: [0,1] &\rightarrow& M \nnn
s & \mapsto & x^\mu (s)
\ee
The beginning of the curve is $\gamma(0)$ and the end of the curve is $\gamma(1)$. Note that in the mathematical literature holonomies are often defined on closed loops, but here we are considering general curves that may or may not be closed.
The holonomy along this path is defined as:
\be
h_\gamma[\bm{A}] &:=&  \Pexp \int_0^1 \rd s \ \dot{\gamma}(s)^\mu A(s)^i_\mu \bm{\tau^i} \equiv \Pexp \int_{\gamma} \bm{A} \nnn
\label{holonomy}
&:=& \sum_{n=0}^\infty \int_0^1 \rd s_1 \int_0^{s_1} \rd s_2 \cdots \int_0^{s_{n-1}} \rd s_n \bm{A}\left(\gamma(s_n)\right) \cdots \bm{A}\left(\gamma(s_1)\right),
\ee
where $\dot{\gamma}(s)^\mu = \frac{\partial \gamma(s)^\mu}{\partial s}$ is a vector tangent to the curve.
Under $\SU(2)$-gauge transformations, the holonomy transforms as:
\be
\label{hgtf}
h_\gamma \rightarrow g_{\gamma(0)} h_\gamma g_{\gamma(1)}^{-1},
\ee
where $g_x$ is an $\SU(2)$-valued function of $x$.

Path-ordering is required in the definition since the connection generally does not commute with itself at different points. This means that two choices of path, say $\gamma$ and $\gamma^\prime$, will generally lead to different results ($h_\gamma \ne h_{\gamma^\prime}$) even if the endpoints remain the same ($\gamma(0) = \gamma^\prime(0)$ and $\gamma(1) = \gamma^\prime (1)$). However, when the curvature is zero ($\bm{F}(\bm{A}) = 0$) the holonomy depends upon its endpoints only:
\be
h_\gamma = h_{\gamma^\prime} = g_{\gamma(0)} g_{\gamma(1)}^{-1},
\ee
so long as $\gamma$ and $\gamma^\prime$ are not closed loops, and are in the same homotopy class, i.e. one curve can be deformed smoothly into the other without crossing any topological defects in $M$ such as particle worldlines.

Now that we have established the necessary properties of holonomies, let us look at the case of a holonomy which follows a path around a particle. Consider a single particle with mass $m$ in the spacetime $M$, at rest at the origin of a cylindrical coordinate system $(t, r, \phi)$. The metric for this spacetime is given by \cite{DJH}:
\be
\rd s^2 = \rd t^2 + \rd r^2 + \left( 1-\frac{m}{2 \pi} \right) \rd \phi^2.
\ee
This can be related to (the Riemannian analog of) a Minkowski spacetime through the transformation $\theta = (1-m/2\pi) \phi$. While the coordinate $\phi$ has the identification $\phi = \phi + 2 \pi$, the Minkowski coordinate has the identification $\theta = \theta +  2\pi - m$. This implies that the metric on a two-surface which intersects the worldline transversely has the geometry of a cone with a deficit angle given by the particle mass.

A frame-field\footnote{The metric is given in terms of the frame-field by $e^i_\mu e^i_\nu$.} and connection describing the above metric is given by \cite{FL}:
\be
\bm{e} &=& \bm{\tau}^0 \rd t +  \left( \cos \phi \bm{\tau}^1 + \sin \phi \bm{\tau}^2 \right) \rd r +  r \left(1-\frac{m}{2\pi}\right) \left( \cos \phi \bm{\tau}^2 - \sin \phi \bm{\tau}^1 \right) \rd \phi, \\
\bm{A} &=& -\frac{m}{2 \pi} \bm{\tau}^0 \rd \phi = -\frac{1}{2 \pi} \bm{p} \rd \phi.
\ee
where in writing the connection we used that $\bm{u} = \bm{\tau}^0$ is the unit vector pointing in the direction of momentum to obtain $\bm{p} = m \bm{\tau}^0$.
One can check that these fields satisfy $\bm{F} = \bm{G} = 0$, remembering that we have excised the particle world line\footnote{There is a delta function contribution to the curvature if we do not excise the worldline \cite{FL}.}.
From now on, we take the path $\gamma$ to be a circular loop around the particle worldline at some fixed values of $r$ and $t$, with a base-point at $b = \gamma(0) = \gamma(1)$. The holonomy $h_{\gamma,b}$ is easy to calculate since the frame-field commutes with itself making path-ordering irrelevant:
\be
\label{hex1}
h_{\gamma,b} = \exp - \int_0^{2 \pi} \frac{1}{2 \pi} \bm{p} \rd \phi = e^{ -\bm{p}}.
\ee
Here the result does not depend on the base-point, but we include it in the notation since the general result will depend on $b$.

Since the connection is flat outside of the particle worldline, this result is the same for any deformation of $\gamma$ that leaves the base-point fixed. To find the holonomy around the particle boundary, we define a path which begins at $b$, goes along a line $\pi$ of constant $\phi$ until it reaches the boundary at point $b^\prime$, circles the boundary $\B$ once, then returns from $b^\prime$ along $\pi$ back to the base-point $b$. This path is a deformation of $\gamma$ which leaves the endpoints fixed, so we have:
\be
h_{\gamma, b} = h_{\pi}(b,b^\prime) h_{\B, b^\prime} h_{\pi}(b, b^\prime)^{-1}.
\ee
Since the connection does not depend on the radial coordinate we have $h_\pi(b,b^\prime) = \mathbb{1}$, and using (\ref{hex1}) we have the holonomy around the particle boundary:
\be
h_{\B,b^\prime} = e^{-\bm{p}}.
\ee

We have so far considered a particle at rest in the frame defined by the $\bm{\tau}^i$ basis. We can repeat the calculation for a particle traveling in an arbitrary timelike\footnote{Nothing in the Riemannian theory is fixing the worldlines to be timelike. However, since the purpose here is to mimic the Lorentzian case we shall adopt these notions.} direction by rotating the direction vector $\bm{u} \rightarrow \tilde{\bm{u}} = g \bm{\tau}^0 g^{-1}$ with an element $g \in \SU(2)$. The connection and frame field can then be written in terms of a new basis $\tilde{\bm{\tau}}^i = g \bm{\tau}^i g^{-1}$ as:
\be
\bm{e} &=&\tilde{\bm{\tau}}^0\rd \tilde{t} +  \left( \cos \tilde{\phi} \tilde{\bm{\tau}}^1+ \sin \tilde{\phi} \tilde{\bm{\tau}}^2\right) \rd \tilde{r} +  \tilde{r} \left(1-\frac{m}{2\pi}\right) \left( \cos \tilde{\phi} \tilde{\bm{\tau}}^2 - \sin \tilde{\phi} \tilde{\bm{\tau}}^1\right) \rd \tilde{\phi}, \\
\bm{A} &=& -\frac{m}{2 \pi}\tilde{\bm{\tau}}^0\rd \tilde{\phi} = -\frac{1}{2 \pi} \tilde{\bm{p}} \rd \tilde{\phi}.
\ee
The coordinate $\tilde{t}$ is associated to the $\tilde{\bm{u}}$ direction, and $(\tilde{r},\tilde{\phi})$ are polar coordinates for any plane running perpendicular to this. The holonomy associated to a circular loop around the particle $\tilde{\gamma}$ at fixed $(\tilde{t}, \tilde{r})$ is given by:
\be
h_{\tilde{\gamma}, \tilde{b}} = e^{-\tilde{\bm{p}}}.
\ee
We have defined our spatial hypersurfaces $\Sigma$ to be spanned by $\bm{\tau}^1$ and $\bm{\tau}^2$, so the loop $\tilde{\gamma}$ is not contained within this plane, while the particle boundary $\B$ is within the $(\bm{\tau}^1 ,\bm{\tau}^2)$-plane. However, since the connection does not depend on the radial or time coordinates, we can deform the curve $\tilde{\gamma}$ while keeping $b$ fixed in a similar manner as done above to find that $h_{\B, \tilde{b}^\prime} = e^{-\tilde{\bm{p}}}$.

There is one further generalization required before we achieve our desired result. We have been using a connection that commutes with itself, but in general the connection may take on different $\su(2)$-values around the loop encircling the particle. Such fields are related to the above via $\SU(2)$-gauge transformations, which we recall here:
\be
\bm{A} \rightarrow g \bm{A} g^{-1} + g \rd g^{-1}, \hspace{1in} \bm{e} \rightarrow g \bm{e} g^{-1} ,
\ee
for an element $g_x \in \SU(2)$. So, any choice of $\SU(2)$-valued field $g_x$ will produce a new frame-field and connection also providing a geometry associated to a point particle. Suppose we have calculated in a particular $\SU(2)$ gauge that $h_{\B, b} = e^{-\bm{p}}$. Under an $\SU(2)$-gauge transformation (\ref{hgtf}) we obtain:
\be
\label{baseh}
h_{\B, b} \rightarrow g_b e^{-\bm{p}} g_b^{-1}.
\ee
The dependence on the choice of gauge and the base-point $b$ is now apparent. One must have knowledge of the particle momentum and the value of the gauge field at the base-point in order to fully determine the holonomy.

Now, the exponential of the momentum $\bm{p} = m \bm{u}$ is a rotation by an angle $m$ about the axis $\bm{u}$, and can be written as:
\be
e^{-\bm{p}} = \mathbb{1} \cos \frac{m}{2} - 2 \bm{u} \sin \frac{m}{2}.
\ee
Using this we can also write the general form of a holonomy around the boundary $\B$ as:
\be
\label{genh2}
h_{\B, b} = \mathbb{1} \cos \frac{m}{2} - 2 \bm{u}_b \sin \frac{m}{2},
\ee
where $\bm{u}_b \equiv g_b \bm{u} g_b^{-1}$ is the axis of rotation as seen from the base-point $b$. In a relativistic theory it is necessary to define a frame of reference in defining the parallel transport $h_{\B, b}$, and this is entering in as a an axis of rotation which depends upon a choice of base-point.
This form of the holonomy is useful in the main article.

\section{Details in calculating the Poisson bracket between the Gauss and mass shell constraints}
\label{aB}
In this appendix we give a detailed calculation of the Poisson bracket $\left\{\mathcal{G}(\L), \psi \right\}$ between the mass shell and Gauss constraints.
For notational convenience we consider the case of a single particle and drop the $\v$ subscript. The generalization to many particles follows simply.

From the constraint definitions (\ref{Gauss}, \ref{mass shell}) we write:
\be
\left\{\mathcal{G}(\L), \psi \right\} = \left\{\int_\Sigma  \lambda^i \left( \rd e^i + \epsilon^{ijk} A^j \wedge e^k \right), 2 \cos^{-1} \frac{W}{2} - m \right\}.
\ee
Integrating by parts on the left side of the bracket, and taking the derivative with respect to $W$ on the right side of the bracket, we obtain:
\be
\left\{\mathcal{G}(\L), \psi \right\} = \left\{ - \oint_\B \lambda^i e^i - \int_\Sigma e^i \left( \rd \lambda^i + \epsilon^{ijk} A^j \lambda^k \right) , W \right\} \left( \frac{-1}{\sqrt{1-\left(\frac{W}{2}\right)^2}} \right).
\ee
Using index notation, we write:
\be
\label{intermediate1}
\left\{\mathcal{G}(\L), \psi \right\}= \frac{1}{\sin\frac{m}{2}}\left( \int_\B \rd s \lambda^i \dot{\B}^b + \int_\Sigma \rd x^2 \epsilon^{ab} \left( \partial_a \lambda^i + \epsilon^{ijk} A^j_a \lambda^k \right) \right) \left\{ e_b^i, \Tr \ h_{\B,b} \right\}
\ee
where $\dot{\B}^b \equiv \partial \B^b / \partial s$ and we used that $W = \Tr \ h_{\B,b} = 2 \cos (m/2)$ as shown in (\ref{traceh}).

In order to evaluate the Poisson bracket between the frame-field and the Wilson loop, we will need to know the bracket between the frame field and the holonomy counter-clockwise around $\B$ with base-point $b$. We write this holonomy as:
\be
h_{\B,b} = \Pexp \oint_\B \rd s \dot{\B}(s)^c A(s)_c^i \bm{\tau}^i,
\ee
The Poisson bracket of $\bm{e}(x)$ with $\bm{A}(y)$ is non-zero only where $x=y$. This splits the holonomy into two:
\be
\label{ehPB}
\left\{ e(x)_b^i, h_{\B, b} \right\} = h_\B(b,x) \bm{\tau}^i h_\B(x,b) \oint_\B \rd s \epsilon_{bc} \dot{\B}(s)^c \delta^2\left( \B (s), x \right),
\ee
where $h_\B(b,x)$ is the holonomy along $\B$ from the base point $b$ to the point $x$, and $h_\B(x,b)$ is the holonomy along the remainder of the loop, from the point $x$ to the base point $b$\footnote{In the case of $x \rightarrow b$ there is ambiguity in this Poisson bracket. As $x$ approaches the base-point at the beginning of the loop we have:
\[
\lim_{x \rightarrow b^+}h_\B(b,x) = \mathbb{1}, \hspace{1in} \lim_{x \rightarrow b^+}h_\B(x,b) = h_{\B,b}^{-1} .
\]
For the limit taken from the other direction we have:
\[
\lim_{x \rightarrow b^-}h_\B(b,x) = h_{\B,b}, \hspace{1in} \lim_{x \rightarrow b^-}h_\B(x,b) = \mathbb{1} .
\]
By convention we make the first choice. Note that either choice gives the same result under a trace.
}.

We substitute this into (\ref{intermediate1}) to obtain:
\be
\left\{\mathcal{G}(\L), \psi \right\} = \frac{1}{\sin\frac{m}{2}}\left( \oint_\B \rd s \lambda^i \dot{\B}^b + \int_\Sigma \rd x^2 \epsilon^{ab} \left( \partial_a \lambda^i + \epsilon^{ijk} A^j_a \lambda^k \right) \right) \Tr (h_{\B,x} \bm{\tau}^i) \oint_\B \rd s \epsilon_{bc} \dot{\B}(s)^c \delta^2\left( \B (s), x \right) ,
\ee
where we used the cyclic property of the trace, and that $h_\B(x,b) h_\B(b,x) = h_{\B,x}$ is a holonomy around the loop $\B$ with base-point $x$ at the point of integration. The first term is zero by symmetry, and using (\ref{genh2}) we finally obtain:
\be
\left\{\mathcal{G}(\L), \psi \right\} = \oint_\B (u_x)^i (\rd_A \L)^i \ ,
\ee
where $\u_x$ is the axis of rotation as seen from the point of integration $x$.
This equation is needed for determining the constraint algebra of the Hamiltonian system.

\end{document}